\documentclass[10pt]{article}
\usepackage{fullpage}
\usepackage{doublespace}
\usepackage{times}
\usepackage[abbr]{harvard}
\usepackage{graphicx}
\usepackage{amssymb,amsmath,amstext}

\newcommand{\Pin}{\parindent 1.5em}

\newcommand{\Te}{\text{eff}}
\newcommand{\Tv}{\text{V}}
\newcommand{\Ts}{\text{S}}

\newcommand{\Avg}[1]{\ensuremath{\left<#1\right>}}
\newcommand{\VAvg}[1]{\ensuremath{\Avg{#1}_{\Tv}}}
\newcommand{\SAvg}[1]{\ensuremath{\Avg{#1}_{\Ts}}}
\newcommand{\Ceff}[1]{\ensuremath{C^{\Te}_{#1}}}

\newcommand{\Sigx}{\sigma_{11}}
\newcommand{\Sigy}{\sigma_{22}}
\newcommand{\Sigz}{\sigma_{33}}
\newcommand{\Tauxy}{\sigma_{12}}

\newcommand{\Sigxy}{\sigma_{12}}
\newcommand{\Sigyz}{\sigma_{23}}
\newcommand{\Sigzx}{\sigma_{31}}
\newcommand{\Epsx}{\epsilon_{11}}
\newcommand{\Epsy}{\epsilon_{22}}
\newcommand{\Epsz}{\epsilon_{33}}
\newcommand{\Gamxy}{\epsilon_{12}}

\newcommand{\ASigx}{\VAvg{\Sigx}}
\newcommand{\ASigy}{\VAvg{\Sigy}}
\newcommand{\ASigz}{\VAvg{\Sigz}}
\newcommand{\ATauxy}{\VAvg{\Tauxy}}

\newcommand{\ASigxy}{\VAvg{\Sigxy}}
\newcommand{\ASigyz}{\VAvg{\Sigyz}}
\newcommand{\ASigzx}{\VAvg{\Sigzx}}
\newcommand{\AEpsx}{\VAvg{\Epsx}}
\newcommand{\AEpsy}{\VAvg{\Epsy}}
\newcommand{\AEpsz}{\VAvg{\Epsz}}
\newcommand{\AGamxy}{\VAvg{\Gamxy}}

\newcommand{\Ca}{\Ceff{11}}
\newcommand{\Cb}{\Ceff{12}}
\newcommand{\Cc}{\Ceff{66}}
\newcommand{\Cd}{\Ceff{22}}
\newcommand{\Xx}{\ensuremath{x^{(\alpha)}_{1}}}
\newcommand{\Yy}{\ensuremath{x^{(\beta)}_{2}}}
\newcommand{\Zz}{\ensuremath{x^{(\gamma)}_{3}}}

\newcommand{\LocXYZ}{\Xx,\Yy,\Zz}
\newcommand{\GloXYZ}{\ensuremath{X_1,X_2,X_3}}
\newcommand{\Vabg}[1]
   {\ensuremath{\frac{1}{8h^3}\int_{-h}^h\int^h_{-h}\int_{-h}^h} {#1}~d\Xx d\Yy d\Zz }

\newcommand{\Abg}{\ensuremath{(\alpha \beta \gamma)}}

\newcommand{\Epsxy}{\VAvg{\epsilon_{12}}}
\newcommand{\Epsyz}{\VAvg{\epsilon_{23}}}
\newcommand{\Epszx}{\VAvg{\epsilon_{31}}}
\newcommand{\Epsxz}{\VAvg{\epsilon_{13}}}

\newcommand{\Suma}{\ensuremath{\sum^{N}_{\alpha=1}}}
\newcommand{\Sumb}{\ensuremath{\sum^{N}_{\beta=1}}}
\newcommand{\Sumg}{\ensuremath{\sum^{N}_{\gamma=1}}}

\begin{document}

  \begin{titlepage}
     \begin{center}
        {\LARGE On the generalized method of cells and the prediction of \protect \\
                effective elastic properties of polymer bonded explosives}\\
        \vspace{48pt} 
        {\large Biswajit Banerjee 
                \footnote{Corresponding author. E-mail: banerjee@eng.utah.edu~~ 
                          Phone: (801)-585-5239~~ Fax: (801)-585-9826}
                ~~and~~Daniel O. Adams}\\
        \vspace{24pt} 
        Dept. of Mechanical Engineering, University of Utah, \\
	  50 S Central Campus Drive \# 2202, Salt Lake City, UT 84112, USA.\\
     \end{center}
  \end{titlepage}

  \section*{Abstract}
  The prediction of the effective elastic properties of polymer bonded explosives using direct numerical simulations is computationally expensive because of the high volume fraction of particles in these particulate composites ($\sim$0.90) and the strong modulus contrast between the particles and the binder ($\sim$20,000).  The generalized method of cells (GMC) is an alternative to direct numerical simulations for the determination of effective elastic properties of composites.  GMC has been shown to be more computationally efficient than finite element analysis based approaches for a range of composites.  In this investigation, the applicability of GMC to the determination of effective elastic properties of polymer bonded explosives is explored.  GMC is shown to generate excellent estimates of effective moduli for composites containing square arrays of disks at volume fractions less than 0.60 and a modulus contrast of approximately 100.  However, for high volume fraction and strong modulus contrast polymer bonded explosives such as PBX 9501, the elastic properties predicted by GMC are found to be considerably lower than finite element based estimates and experimental data.  Simulations of model microstructures are performed to show that normal stiffnesses are underestimated by GMC when stress-bridging due to contact between particles is dominant.  Additionally, the computational efficiency of GMC decreases rapidly with an increase in the number of subcells used to discretize a representative volume element.  The results presented in this work suggest that GMC may not be suitable for calculating the effective elastic properties of high volume fraction and strong modulus contrast particulate composites.  Finally, a real-space renormalization group approach called the recursive cell method (RCM) is explored as an alternative to GMC and shown to provide improved estimates of the effective properties of models of polymer bonded explosives. 

  \vspace{12pt}
  {\bf Keywords} :  Effective Properties, Particulate Composites, High Volume
                        Fraction, Strong Modulus Contrast, Stress-Bridging,
                        Method of Cells, Real-Space Renormalization Group 

  \Pin
  \section{Introduction}
  The generalized method of cells (GMC) ~\cite{Aboudi96_1,Paley92} is a semi-analytical method of determining the effective properties of composites.  In this method, a representative volume element (RVE) of the composite under consideration is discretized into a regular grid of subcells.  Equilibrium and compatibility are satisfied on an average basis across subcells using integrals over subcell boundaries.  GMC generates a matrix of algebraic expressions containing information about subcell material properties.  The effective stiffness of the composite can be obtained by inverting this matrix.

  One advantage of GMC over other numerical techniques is that the full set of effective elastic properties of a composite can be calculated in one step instead of solving a number of boundary value problems with different boundary conditions.  GMC has also been found to be more computationally efficient that finite element calculations for fiber reinforced composites~\cite{Aboudi96_1,Wilt95}, since far fewer GMC subcells than finite elements are necessary to obtain the same degree of accuracy.  The problem of discretization is also minimized since a regular rectangular grid is used in GMC.  

  The generalized method of cells is discussed briefly in this work.  Effective stiffnesses predicted by this method are compared with accurate numerical predictions for square arrays of disks~\cite{Greeng98}.  The method is then applied to two-dimensional models of a general polymer bonded explosive and to the microstructure of PBX 9501 using a two-step procedure similar to that of Low et al.~\cite{Low94}.  GMC estimates of elastic properties are compared with predictions from detailed finite element calculations.  The performance of GMC is explored for several microstructures with contacting particles and some shortcomings of the method are identified.  Finally, the recursive cell method (RCM) is explored as an alternative to direct GMC calculations in the prediction of effective properties of polymer bonded explosives.

  \section{The generalized method of cells}
  Figure~\ref{fig:GMCRVE} shows a schematic of the RVE, the subcells and the notation~\cite{Aboudi91} used in GMC.  In the figure, (\GloXYZ) is the global coordinate system of the RVE and (\LocXYZ) is the coordinate system local to a subcell denoted by $\Abg$.
  It is assumed that the displacement function $u^{\Abg}_{i}$ varies linearly within a subcell~\Abg~and can be written in the form
  \begin{equation}
     u^{\Abg}_{i}(\LocXYZ) = w^{\Abg}_{i}(\GloXYZ) + 
               \Phi^{\Abg}_{i}\Xx + \Theta^{\Abg}_{i}\Yy + \Psi^{\Abg}_{i}\Zz 
	\label{E:DispField}
  \end{equation}
  where $i$~represents the coordinate direction and takes the values 1, 2 or 3; $w^{\Abg}_{i}$~is the mean displacement at the center of the subcell~\Abg; and $\Phi^{\Abg}_{i}$, $\Theta^{\Abg}_{i}$, and $\Psi^{\Abg}_{i}$ are constants local to the subcell that represent gradients of displacement across the subcell.  The strain-displacement relations for the subcell are given by 
  \begin{equation}
    \epsilon^{\Abg}_{ij}= \frac{1}{2}(\partial_{i}u^{\Abg}_{j}+
		    \partial_{j}u^{\Abg}_{i})
  \end{equation}
  where $\partial_1 = \partial/\partial\Xx$, $\partial_2 =\partial/\partial\Yy$, and $\partial_3 = \partial/\partial\Zz$.  Since polymer bonded explosives are isotropic particulate composites, the following brief description of GMC assumes that the RVE is cubic and all subcells are of equal size.  If each subcell $\Abg$ has the same dimensions $(2h,2h,2h)$ then the average strain in the subcell is defined as a volume average of the strain field over the subcell as
  \begin{equation}
     \SAvg{\epsilon^{\Abg}_{ij}} = \Vabg{\epsilon^{\Abg}_{ij}} 
     \label{eq:volAvStrain}
  \end{equation}
  The average strain in the subcell can be obtained in terms of the displacement field variables.  It is assumed that there is continuity of traction at the interface of two subcells.  The displacements and tractions are assumed to be periodic at the boundaries the RVE.  Applying the displacement continuity equations on an average basis over the interfaces between subcells, the average strain in the RVE can be expressed in terms of the subcell strains.  The average subcell stresses can be obtained from the subcell strains using the traction continuity condition and the stress-strain relations of the materials in the subcells.  A relationship between the subcell stresses and the average strains in the RVE is thus obtained.

For orthotropic, transversely isotropic or isotropic materials, the approach discussed above leads to the decoupling of the normal and shear response of the RVE.  This decoupling leads to two systems of equations relating the subcell stresses and the average strains in the RVE.  For the normal components of strain, the system of equations can be written as
  \begin{equation} \label{E:NormalEquation}
     \begin{bmatrix}
     \mathbf{M1}_{1} & \mathbf{M1}_{2} & \mathbf{M1}_{3} \\
     \mathbf{M2}_{1} & \mathbf{M2}_{2} & \mathbf{M2}_{3} \\
     \mathbf{M3}_{1} & \mathbf{M3}_{2} & \mathbf{M3}_{3} 
     \end{bmatrix}
     \begin{bmatrix}
     \mathbf{T}_{1} \\ \mathbf{T}_{2} \\ \mathbf{T}_{3}
     \end{bmatrix} 
     = 2N
     \begin{bmatrix}
        \mathbf{H} \\ \mathbf{0} \\ \mathbf{0} 
     \end{bmatrix}
     \AEpsx + 2N
     \begin{bmatrix}
        \mathbf{0} \\ \mathbf{H} \\ \mathbf{0} 
     \end{bmatrix}
     \AEpsy + 2N
     \begin{bmatrix}
        \mathbf{0} \\ \mathbf{0} \\ \mathbf{H} 
     \end{bmatrix}
     \AEpsz 
  \end{equation}
  where $N$ is the number of subcells per side of the RVE.
  The corresponding system of equations for the shear components is of the form
  \begin{equation} \label{E:ShearEquation}
   \begin{bmatrix}
	\mathbf{M4} & \mathbf{0} & \mathbf{0} \\
	\mathbf{0} & \mathbf{M5} & \mathbf{0} \\
	\mathbf{0} & \mathbf{0} & \mathbf{M6} 
   \end{bmatrix}
   \begin{bmatrix}
	\mathbf{T}_{12} \\ \mathbf{T}_{23} \\ \mathbf{T}_{13}
   \end{bmatrix}
   = 2N^2
   \begin{bmatrix}
	\mathbf{H} \\ \mathbf{0} \\ \mathbf{0}
   \end{bmatrix}
   \Epsxy + 2N^2
   \begin{bmatrix}
	\mathbf{0} \\ \mathbf{H} \\ \mathbf{0}
   \end{bmatrix}
   \Epsyz + 2N^2
   \begin{bmatrix}
	\mathbf{0} \\ \mathbf{0} \\ \mathbf{H}
   \end{bmatrix}
   \Epsxz 
  \end{equation}
  In equations~(\ref{E:NormalEquation}) and (\ref{E:ShearEquation}) the $\mathbf{M}$ matrices contain material compliance terms.  The $\mathbf{T}$ matrices contain the average subcell stresses.  The vector $\mathbf{H}$ contains the dimensions of the subcells.  Thus, a sparse system of equations of size 3$N^2$ is produced that relates the subcell stresses to the average strains in the RVE.  After inverting these equations and with some algebraic manipulation, explicit algebraic expressions for the individual terms of the effective stiffness matrix can be obtained.  These stress-strain equations that relate the average RVE stresses to the average RVE strains are of the form
  \begin{equation}
    \begin{bmatrix}\ASigx\\\ASigy\\\ASigz\\\ASigyz\\\ASigzx\\\ASigxy\end{bmatrix}  =
    \begin{bmatrix} \Ceff{11} & \Ceff{12} & \Ceff{13} & 0 & 0 & 0\\
                    \Ceff{12} & \Ceff{22} & \Ceff{23} & 0 & 0 & 0\\
                    \Ceff{13} & \Ceff{23} & \Ceff{33} & 0 & 0 & 0\\
                    0 & 0 & 0 & \Ceff{44} & 0 & 0 \\
                    0 & 0 & 0 & 0 & \Ceff{55} & 0 \\
                    0 & 0 & 0 & 0 & 0 & \Ceff{66} \end{bmatrix}
    \begin{bmatrix}\AEpsx\\\AEpsy\\\AEpsz\\2\Epsyz\\2\Epszx\\2\Epsxy\end{bmatrix}
    \label{eq:sigepsFE1}
  \end{equation}
  where $\Ceff{ij}$ are the terms of the effective stiffness matrix. Details of the algebraic expressions for these terms have been published by other researchers~\cite{Pindera97}.

In GMC, the number of equations to be solved equals the number of subcells raised to the $d$~th power, where $d$ is the number of dimensions in the problem. As a result, the computational efficiency of GMC decreases as the number of subcells increases.  This issue has been partially resolved~\cite{Orozco97} by identifying the sparsity characteristics of the system of equations and by using the Harwell-Boeing suite of sparse solvers.  The computational efficiency of GMC has been further improved after a reformulation~\cite{Pindera97,Bednar97} that takes advantage of the continuity of tractions across subcells to obtain a system of $O(N^2)$ equations in three dimensions.

Due to decoupling of the normal and shear response of the RVE, the shear components of the stiffness matrix obtained from GMC are the harmonic means of the subcell shear stiffnesses and of the form
\begin{equation}
1/\Ceff{66} = 1/N^3\Suma\Sumb\Sumg 1/C^{\Abg}_{66}.
\end{equation}
Bednarcyk and Arnold (2001) suggest that this lack of coupling makes for an ``ultra-efficient'' micromechanics model.  However, this lack of coupling can lead to gross underestimation of shear moduli for high volume fraction and high modulus contrast materials such as polymer bonded explosives.  Recently, researchers~\cite{Williams99,Gan00} have attempted to solve the problem by using higher order expansions for the displacement and by explicitly satisfying both subcell equilibrium and compatibility.  However, these approaches decrease the computational efficiency of GMC considerably and are not explored in this work.

  \section{Validation - square arrays of disks}
  In this section, estimates of effective properties from GMC are compared with accurate numerical results for square arrays of disks.  Square RVEs containing square arrays of disks exhibit square symmetry.  The two-dimensional linear elastic stress-strain relation for these RVEs can be written as
  \begin{equation}
    \label{eq:squareSymm}
    \begin{bmatrix} \ASigx\\\ASigy\\\ATauxy\end{bmatrix}  =
    \begin{bmatrix} K_{\Te}+\mu_{\Te}^{(1)}&K_{\Te}-\mu_{\Te}^{(1)}&0 \\ 
                    K_{\Te}-\mu_{\Te}^{(1)}&K_{\Te}+\mu_{\Te}^{(1)}&0 \\ 
		    0 & 0 & \mu_{\Te}^{(2)} \end{bmatrix}
    \begin{bmatrix} \AEpsx\\\AEpsy\\2\AGamxy\end{bmatrix} 
  \end{equation}
  where $K_{\Te}$ is the two-dimensional effective bulk modulus, $\mu_{\Te}^{(1)}$ is the effective shear modulus when a shear stress is applied along the diagonals of the RVE, and $\mu_{\Te}^{(2)}$ is the effective shear modulus when a shear stress is applied along the edges of the RVE.  These three effective moduli have been determined accurately, using an integral equation approach, by Greengard and Helsing (1998) for square arrays of disks containing a range of disk volume fractions.  

To compare the effective moduli predicted by GMC with those from the integral equation calculations~\cite{Greeng98}, RVEs containing disk volume fractions from 0.10 to 0.70 were created.  These RVEs were discretized into 64$\times$64 equal sized subcells.  The effective stiffness matrix of each RVE was calculated using GMC.  Finally, the two-dimensional effective moduli for each RVE were calculated from the effective stiffness matrix using the relations
  \begin{equation}
    K_{\Te} = 0.5(\Ceff{11} + \Ceff{12})~,~
    \mu^{(1)}_{\Te} = 0.5(\Ceff{11} - \Ceff{12})~,~
    \mu^{(2)}_{\Te} = \Ceff{66}.
  \end{equation}

  Figure~\ref{fig:helGMC} shows the moduli predicted by GMC and those from the integral equation method of Greengard and Helsing (G\&H) for disk volume fractions from 0.10 to 0.70.
The material properties of the disks and the binder used in the calculations are shown in Table~\ref{tab:helprop}.  
  The effective bulk moduli ($K_{\Te}$) and diagonal shear moduli ($\mu_{\Te}^{(1)}$) obtained from the GMC calculations are within 4\% of those obtained by the integral equation method for all volume fractions up to 0.60.  At a volume fraction of 0.70, the GMC predictions for bulk modulus and diagonal shear modulus are 4\% and 11\% less, respectively.  For the shear modulus $\mu_{\Te}^{(2)}$, the GMC predictions are around 4\% to 10\% less than the estimates of Greengard and Helsing for volume fractions from 0.10 to 0.60.  The difference is around 24\% for a volume fraction of 0.70.  

  These results show that GMC estimates are quite accurate for composites containing square arrays of disks with volume fractions up to 0.60, confirming results reported elsewhere~\cite{Aboudi96_1}.  In the next section, GMC is used to determine the effective properties of models of polymer bonded explosives and the results are compared to detailed finite element calculations and experimental data.

\section{Modeling polymer bonded explosives}
  Polymer bonded explosive (PBX) materials typically contain around 90\% by volume of particles surrounded by a binder.  The particles consist of a mixture of coarse and fine grains with the finer grains forming a filler between coarser grains.  Modeling the microstructure of these materials is difficult due to the complex shapes of HMX particles and the large range of particle sizes.  Two-dimensional approximations of the microstructure of PBXs based on digital images~\cite{Benson99} have been used to study some aspects of the micromechanics of PBXs.  However, such microstructures are difficult to generate and require complex image processing techniques and excellent image quality to accurately capture details of the material.  A combination of Monte Carlo and molecular dynamics techniques have also been used to generate three-dimensional models of PBXs~\cite{Baer01}.  Microstructures containing spheres and oriented cubes have been generated using these techniques and appear to represent PBX microstructures well.  However, the generation of microstructures using dynamics-based methods is extremely time consuming when tight particle packing is required, as is the case for volume fractions above 0.70.  

  Comparisons of finite element predictions with exact relations for the effective properties of composites~\cite{Banerjee02th} have shown that detailed finite element estimates can be used as a benchmark to check the accuracy of predictions from GMC.  In this investigation, manually generated PBX microstructures containing symmetrically distributed circular particles are used initially to compare GMC and finite element predictions.  The two-dimensional microstructures contain 90\% particles by volume and use two particle length scales.  Two-dimensional models containing randomly distributed circular particles that reflect the actual particle size distribution of PBX 9501 are next modeled with GMC and the results compared to finite element estimates.  
  
  The material properties used for the particles, the binder, and PBX9501 in these calculations are shown in Table~\ref{tab:exptCPBX}.  These properties correspond to those of HMX (the explosive particles), the binder, and PBX 9501 at 25$^o$ C and a strain rate of 0.05/s~\cite{Wetzel99}.  

  \subsection{Simplified models of PBX materials}
  GMC and finite element calculations were performed for the six, manually generated, simplified model microstructures of polymer bonded explosives shown in Figure~\ref{fig:gmcfemManual}.
  These representative volume elements (RVEs) contain one or a few relatively large particles surrounded by smaller particles to reflect common particle size distributions of PBXs.  The volume fraction of particles in each of these models is around 0.90$\pm$0.005.  The binder material surrounds all particles in the six microstructures.  

  For the GMC calculations, a square grid was overlaid on the RVEs to generate subcells.  Two different approaches were used to assign materials to subcells before the determination of effective properties of the RVE.  In the first approach, referred to as the ``binary subcell approach'', a subcell was assigned the material properties of particles if more than 50\% of the subcell was occupied by particles.  Binder properties were assigned otherwise.  Figure~\ref{fig:pbxManualgmc}(a) shows a schematic of the binary subcell approach.  In the second approach, called the ``effective subcell approach'', a method of cells calculation~\cite{Aboudi91} was used to determine the effective properties of a subcell based on the cumulative volume fraction of particles in the subcell~\cite{Banerjee00}.  Figure~\ref{fig:pbxManualgmc}(b) shows a schematic of the effective subcell approach.  After the subcells were assigned material properties, the GMC technique was used to compute the effective properties of the RVE.  

 Note that the particles are not resolved well when materials are assigned to subcells in this manner if the number of subcells is small.  However, the large size of the matrix to be inverted in GMC limits the number of subcells that can be used to discretize the RVE.   If the binary subcell approach is used to assign subcell materials, contacting particles are created where there are none in the actual microstructure, leading to the prediction of higher than actual stiffness values.  The effective subcell approach improves upon the binary subcell approach by ``smearing''the material properties at the boundaries of particles and thus reducing the particle contact artifacts caused by discretization errors.  

  For validating the GMC results, detailed finite element (FEM) calculations were performed using six-noded triangular elements to accurately model the geometry of the particles.  Around 65,000 nodes were used to discretize each of the models.  The volume average stress and strain in each RVE was determined for applied normal and shear displacements.  Periodicity was enforced through displacement boundary conditions.   Since these finite element calculations serve to validate the GMC calculations, further mesh refinement was explored and the results were found to converge those from the 65,000 node finite element models.

  Table~\ref{tab:pbxManualGMC} lists the effective stiffnesses of the six RVEs shown in Figure~\ref{fig:gmcfemManual} from GMC and FEM calculations.
  On average, the GMC calculations using the binary subcell approach predict values of $\Ca$ that are around 2.5 times the FEM based values.  The values of $\Ca$ from the effective subcell approach based GMC calculations are closer to the FEM estimates than the binary subcell based GMC estimates.  The GMC and FEM estimates of $\Cb$ are quite close.  The values of $\Cc$ from GMC are only 10\% of the FEM values.  The low values of $\Cc$ are obtained because GMC predict effective shear stiffnesses that are harmonic means of subcell shear stiffnesses.  Models 3 and 4 (Figure~\ref{fig:gmcfemManual}) produce agreement in $\Ca$ and $\Cb$ between the binary subcell approach based GMC calculations and FEM (within 5\%) whereas the other four models produce considerable differences in predictions.  These large differences are produced by discretization errors introduced in the GMC approach that lead to continuous stress-bridging paths across the RVEs and hence to increased stiffness. 

 However, if the predicted effective stiffnesses shown in Table~\ref{tab:pbxManualGMC} are compared with the experimental effective stiffness of PBX 9501 (shown in Table~\ref{tab:exptCPBX}), it can be observed that the models predict values of $\Ca$ and $\Cb$ that are around 10\% of the experimental values.  Hence, these simplified models are not appropriate for the modeling of PBX 9501.  The next section explores models based on the actual particle size distribution of PBX 9501.

  \subsection{Models of PBX 9501}\label{sec:pbx9501}
  Coarse and fine particles of HMX are blended in a ratio of 1:3 by weight and compacted in the process of manufacturing PBX 9501.  Figure~\ref{fig:modelPBX}(a) shows four RVEs of PBX 9501 based on the particle size distribution of the dry blend of HMX~\cite{Wetzel99} prior to compaction.  Figure~\ref{fig:modelPBX}(b) shows four RVEs based on the particle size distribution of pressed PBX 9501~\cite{Skid98}.  The larger particles are broken up in the pressing process leading to a larger proportion of smaller particles in pressed PBX 9501.  The models of the dry blend have been labeled ``DB'' while those of pressed PBX 9501 have been labeled ``PP''.

  GMC calculations were performed on the PBX 9501 RVEs after discretizing each RVE into 100$\times$100 subcells and assigning materials to subcells using the effective subcell approach.  In order to validate the GMC predictions, FEM calculations were also performed on the RVEs after discretizing each RVE into 350$\times$350 four-noded square elements.  The binary subcell approach was applied to assign materials to elements for the FEM calculations.  

  The particles in each RVE were assigned properties of HMX from Table~\ref{tab:exptCPBX}.  However, since particles occupy 92\% of the total volume in actual PBX 9501 while the sample microstructures could be filled only up to $\sim$86\%, an intermediate homogenization step was required to determine the properties of the binder.  To produce the desired 92\% volume of particles, a fine-particle filled binder containing 36\% particles by volume, or ``dirty'' binder, was assumed.  The effective elastic properties of the dirty binder were calculated using the differential effective medium approximation~\cite{Markov00}.  

  Table~\ref{tab:pbxFEMGMC} shows the effective stiffness from FEM and GMC calculations for the models of PBX 9501 shown in Figure~\ref{fig:modelPBX}.
  For all microstructures, the values of $\Ca$ and $\Cd$ predicted by GMC are less than 5\% of the FEM values and less than 10\% of the experimental values for PBX 9501 (shown in Table~\ref{tab:exptCPBX}).  The FEM estimates increase with RVE size (varying from 150\% to 450\% of the experimental values), reflecting the dependence of predicted stiffnesses on microstructure, discretization and particle size distribution.  The RVEs contain numerous particle to particle contacts, the number of which increases with increase in RVE size.  These contacts lead to significant stress-bridging and hence relatively high values of stiffness as is reflected in the FEM predictions.  However, stress-bridging is not incorporated accurately in the GMC approach leading to considerably lower values of effective stiffness.  The issue of stress-bridging is further explored in the following section.

The values of $\Cc$ predicted by GMC are around 0.5\% of the FEM predictions.  The reason for this large difference is that the effective shear stiffness predicted by GMC is simply the harmonic mean of the subcell shear moduli and only provides a lower bound on the shear stiffness.  

  \section{Stress-Bridging}
  Comparisons of effective stiffness properties predicted by GMC with other numerical estimates have shown that GMC performs quite well for low modulus contrast materials with volume fractions below 60\%.  However, for high modulus contrast materials with high particle volume fractions, GMC usually predicts considerably lower effective stiffnesses than finite element calculations.  In this section, GMC is applied to selected microstructures containing stress-bridging and the predicted properties are compared to finite element estimates.  The goal is to demonstrate that the effects of stress-bridging on effective properties are inaccurately described by GMC.  

  \subsection{Corner bridging : X-shaped microstructure}
  In the RVE shown in Figure~\ref{fig:bridge_1}, the particles are square, arranged in the form of an `X', and occupy a volume fraction of 25\%. The particles transfer stress through corner contacts.  
  The effective properties of the X-shaped microstructure shown in Figure~\ref{fig:bridge_1} were calculated using the properties of HMX and five different binders with Young's moduli that range from 0.7 MPa to 7000 MPa, as shown in Table~\ref{tab:bridge1atoe}.  

  Figure~\ref{fig:bridge1GMC} shows the variation in the effective stiffness properties $\Ceff{11}$ and $\Ceff{66}$ of the X-shaped microstructure with increasing Young's modulus contrast between the particles and the binder ($E_p/E_b$).  
These effective stiffness properties have been calculated using both finite elements (FEM) (256$\times$256 elements) and GMC (64$\times$64 subcells).  The FEM and GMC estimates are in good agreement for Young's modulus contrasts of 200 or less.  For higher Young's modulus contrasts, the effective stiffness properties predicted by GMC are much lower than those predicted by FEM.  Note that the FEM estimates do not change significantly with increased discretization, implying that the solution has converged.  The effect of corner singularities is also averaged out while calculating the effective properties using FEM.  If it is assumed that the FEM estimates are close to the actual effective moduli of the RVE, the GMC estimates for high modulus contrasts are orders of magnitude lower than the actual effective moduli.  Hence, GMC does not capture the stiffening effect of corner contacts accurately.

  Since the corner stress-bridging problem involves high stress concentrations that are not resolved well by finite elements, it is possible that the FEM calculations overestimate the effective properties of the X-shaped microstructure.  Such corner singularities are minimized in the microstructures studied in the next section where the effect of stress-bridging along particles edges is studied. 

  \subsection{Edge bridging}
  Figure~\ref{fig:bridge2GMC} shows five RVEs (A through E) in which the degree of stress-bridging is increased progressively from corner bridging, to partial edge bridging, and finally to continuous stress-bridging across the RVE.  In Figure~\ref{fig:bridge2GMC}, the `1' direction corresponds to the $x$-axis and the `2' direction corresponds to the $y$-axis.

  Model A contains a square particle that occupies 25\% of the volume, is centered in the RVE, and does not have any stress-bridging.  Model B contains three particles that contact along a diagonal of the RVE.  In model C, particle contact is increased to produce a single line of stress-bridging in the $x$-direction along the center of the RVE.  Model D extends the line of contact in the $x$-direction to an area of contact in the $x$-direction.  In Model E, particle bridging across the RVE is extended to both directions. The material properties of the constituents of PBX 9501 at room temperature and low strain rate were used for the calculations (Table~\ref{tab:exptCPBX}).  GMC simulations of the RVEs were performed using 100$\times$100 subcells while the validating finite element calculations were performed using approximately 10,000 eight-noded quadrilateral elements.  Table~\ref{tab:bridge2GMC} shows the effective stiffnesses of the five models obtained from GMC and finite element (FEM) calculations.

  As expected in model A, GMC and FEM predict nearly the same values of effective stiffness since there is no stress-bridging in the model (the effective stiffness is determined primarily by the volume fraction occupied by the square particle).  However, FEM calculations for model B show that the diagonal stress-bridge in the model produces a higher stiffness than would occur if only the volume fraction occupied by the particles were considered in the calculation of effective stiffness.  

  The GMC calculations for model B predict values of $\Ca$ and $\Cb$ that are lower than the FEM estimates by a factor of 18.  This discrepancy implies that the diagonal stress-bridge in model B is not detected by the GMC calculations.  The value of $\Cc$ from FEM is around 1,400 times that from GMC.  This difference shows that, in the presence of stress-bridging, the shear stiffness can be considerably underestimated by GMC, even for low volume fraction composites.   

  Model C has a continuous path through particles along the $x$-axis (the `1' direction) and another continuous particle path along one diagonal.  Intuitively, the stress-bridge path along the `1' direction is expected to primarily affect the normal components of stiffness ($\Ca$, $\Cb$, $\Cd$) while particle contact along the diagonal is expected to affect the shear stiffness ($\Cc$).  These paths are shown by dashed lines (for normal stress-bridging) and by dotted lines (for shear stress-bridging) in Figure~\ref{fig:bridge_2C}.  
  Results for model C in Table~\ref{tab:bridge2GMC} show that FEM predicts a considerable stiffening in the `1' direction while GMC does not appear to account for these stress-bridges.  Since the shear stiffness from GMC is simply a harmonic means of the subcell stiffnesses, $\Cc$ is not affected at all by geometry and only increases in proportion with the volume fraction of particles in the RVE.  

  The estimates of $\Ca$ for model D and of $\Ca$, $\Cb$, $\Cd$ for model E show that GMC can capture the effect of stress-bridging, provided there are continuous rows of particles with edge-to-edge contacts extending completely across the RVE.

  These studies of stress-bridging explain why GMC underestimates the effective modulus of the PBX 9501 models shown in Figure~\ref{fig:modelPBX}.  In all these models, if 100$\times$100 subcells are used to discretize the RVE, there are no rows or columns of subcells extending across the RVE that contain no binder.  Though corner contacts and other continuous stress paths exist in the PBX 9501 models, the effects of these stress-bridging paths are not incorporated into the GMC estimates of effective stiffness.  The strain-compatible or shear-coupled method of cells~\cite{Williams99,Gan00} approaches may be able to overcome some of these deficiencies of GMC.  However, the computational efficiency of GMC is greatly reduced when these modifications are incorporated into GMC and hence the attractiveness of this micromechanics approach as an alternative to finite element analysis is also reduced. 
  
  \section{The recursive cell method}
  The recursive cell method (RCM)~\cite{Banerjee02th} is a real-space renormalization group~\cite{Wilson71,Wilson79} approach for calculating the effective elastic properties of composites that has been developed to address the shortcomings of GMC while retaining high computational efficiency.  A schematic of RCM is shown in Figure~\ref{fig:RCM}.  
  In RCM, as in GMC, the RVE is first discretized into a regular grid of subcells.  For the first iteration of the recursive process, the subcells are assigned material properties based on the particle distribution in the RVE using the binary subcell approach discussed earlier.  The subcells in the original grid are then grouped into blocks of n$\times$n subcells.  The effective elastic stiffness matrix of each of the blocks is calculated using a suitable homogenization approach such as GMC or FEM.  Effective stiffnesses are assigned to each block, resulting in a new, coarser grid.  This procedure is repeated until only one homogeneous block remains.  The properties of this homogeneous block are the effective properties of the RVE.

  Studies on the recursive cell method~\cite{Banerjee02th} have shown that the method leads to an upper bound on the effective elastic properties if a FEM approach is used to homogenize blocks of subcells.  As the number of elements used to discretized a block is increased, the value of the upper bound decreases and a more accurate estimate of the effective properties is obtained.  GMC is an attractive alternative to the FEM approach for homogenization since less discretization is required to arrive at the same level of accuracy.  

  In the previous section, GMC has been shown to not properly account for stress-bridging in the absence of continuous stress-bridge paths across a RVE.  However, error due to improper stress-bridging is reduced when GMC is used as the homogenizer in RCM because the probability of the existence of continuous stress-bridging paths across blocks of subcells is greater than that for the whole RVE.  In addition, homogenization errors due to the overestimation or underestimation of stress-bridging in sections of the RVE are averaged out if the particle distribution is sufficiently random.  

  A second source of error in GMC is the underestimation of the shear stiffness term $\Cc$.  However, this error can be avoided while using RCM to determine the effective elastic properties of PBXs because, for macroscopically isotropic materials such as PBXs, relatively accurate estimates of the effective shear stiffness can be obtained from the effective normal stiffness terms~\cite{Banerjee02th} and therefore direct estimates of $\Cc$ are not required.  On the other hand, if the composite is not macroscopically isotropic, a FEM homogenizer~\cite{Banerjee02th} can be used to determine the value of $\Cc$ instead of GMC.

 The RCM technique has been applied to the four microstructures of the dry blend and pressed PBX 9501 shown in Figure~\ref{fig:modelPBX}(a) and ~\ref{fig:modelPBX}(b), respectively.  Each RVE was discretized into blocks of 256$\times$256 square subcells of equal size.  At each stage of recursion, blocks of 2$\times$2 subcells were homogenized using GMC.  

The values of $\Ceff{11}$ for the four dry blend microstructures obtained from finite element (FEM) calculations, GMC calculations and RCM calculations are compared in Figure~\ref{fig:dryAll}(a).
The RCM estimates of $\Ceff{11}$ for the four microstructures vary from 90\% to 150\% of the FEM estimates. These RCM estimates are a considerable improvement over the GMC estimates shown as black bars in Figure~\ref{fig:dryAll}(a).  Comparisons of $\Ca$ for the four pressed PBX 9501 microstructures (shown in Figure~\ref{fig:modelPBX}(b)) are shown in Figure~\ref{fig:dryAll}(b).  For pressed PBX 9501, the RCM estimates vary between 84\% and 180\% of the finite element estimates.  These RCM estimates are also a considerable improvement over the GMC estimates of effective properties.  RCM estimates of $\Ceff{22}$ and $\Ceff{12}$ for the dry blend and pressed PBX 9501 have also been found to be in much better agreement with FEM results than the GMC estimates.  

As was expected, the estimated value of $\Ceff{66}$ from RCM is quite low compared to both finite element estimates and experimental data.  An improved estimate of $\Ceff{66}$ can be obtained if the shear stiffness of each RCM block is calculated using finite elements~\cite{Banerjee02th}.  The normal stiffnesses $\Ca$, $\Cb$, and $\Cd$ can be still be calculated using GMC, taking advantage of the absence of shear coupling. 

These results show that the RCM approach, in conjunction with a GMC homogenizer, can be used to arrive at reasonably accurate estimates of the effective properties of PBX materials.  The RCM approach can therefore be used as an alternative to direct GMC calculations for high volume fraction, strong modulus contrast materials such as polymer bonded explosives.

  \section{Summary and conclusions}
  The generalized method of cells (GMC) has been found to accurately predict the effective elastic properties of composites containing square arrays of disks for volume fractions up to 0.60.  However, for two-dimensional models of the polymer bonded explosive PBX 9501, estimates of effective elastic properties from GMC have been found to be considerably lower than both experimental values and estimates based on finite element (FEM) calculations.

The lower values of normal stiffness predicted by GMC for PBX 9501 are due to inadequate incorporation of particle stress bridging into the approach.  Model representative volume elements (RVEs) with corner and edge stress bridging show that corner bridging is ignored by GMC while edge stress bridging is incorporated only if continuous stress bridges exist along entire rows or columns of subcells that traverse the length of the RVE.  Low values of effective shear stiffness predicted by GMC can be attributed to the use of a harmonic mean of subcell shear stiffnesses to determine the effective shear stiffness of a RVE.  The harmonic mean is a lower bound on the effective shear stiffness and is not applicable for microstructures where there is significant interaction between particles.  

Improvements suggested to GMC that incorporate normal-shear coupling and strain compatibility across subcells have the potential to overcome some of these weaknesses of GMC.  However, these improvements lead to much larger systems of equations and a considerable increase in the computational cost of the method.  The requirement of inverting a large matrix to obtain the effective properties makes the generalized method of cells very inefficient as the number of subcells increases.  When materials such as PBX 9501 are modeled, the number of subcells needed to represent a random distribution of particles necessarily becomes large. In such situations, the generalized method of cells becomes inefficient and it may be preferable to perform finite element analyses to determine the effective properties.  Thus, GMC does not appear to be an improvement over finite element analyses for high volume fraction, high modulus contrast particulate composites such as polymer bonded explosives.

A computationally efficient alternative to both direct GMC and finite elements is the recursive cell method (RCM) with GMC being used to homogenize blocks of subcells.  RCM estimates of normal stiffness terms for models of PBX 9501 show considerable improvement compared to GMC estimates.  The RCM estimates of shear stiffness can be improved if FEM is used, rather than GMC, to determine the effective shear stiffness of blocks of subcells.  RCM, with a combination of GMC and FEM being used to homogenize blocks of subcells, has the potential of providing fast and accurate estimates of the effective properties of polymer bonded explosives.

  \section*{Acknlowledgements}
  This research was supported by the University of Utah Center for the Simulation of Accidental Fires and Explosions (C-SAFE), funded by Department of Energy grant DE-FG03-02ER45914.  The authors would also like to thank Prof. Graeme Milton and Dr. Brett Bednarcyk for their suggestions.

  \clearpage
  \flushleft
  \protect\begin{flushleft}
  \begin{figure}
 	\caption{RVE, subcells and notation used in GMC.}
        \label{fig:GMCRVE}
  \end{figure}

  \begin{figure}
    \caption{Comparison of effective moduli of square arrays of disks from
	     Greengard and Helsing (1998) (G\&H) and GMC calculations.}
    \label{fig:helGMC}
  \end{figure}

  \begin{figure}
        \caption{Manually generated microstructures containing $\sim$ 90\% circular
                 particles by volume.}
        \label{fig:gmcfemManual}
  \end{figure}

  \begin{figure}
 	\caption{Schematics of the application of the binary subcell approach and the 
	         effective subcell approach in GMC calculations.}
        \label{fig:pbxManualgmc}
  \end{figure}

  \begin{figure}
        \caption{Microstructures containing circular particles based on the
                 particle size distribution of the dry blend (DB) of PBX 9501
                 and of pressed (PP) PBX 9501.}
        \label{fig:modelPBX}
  \end{figure}

  \begin{figure}
 	\caption{RVE used for corner stress-bridging model.}
        \label{fig:bridge_1}
  \end{figure}

  \begin{figure}
 	\caption{Variation of effective stiffness with modulus contrast for `X'-shaped
		 microstructure. The Young's modulus contrast is the ratio of the Young's modulus of the particles to that of the binder.}
        \label{fig:bridge1GMC}
  \end{figure}

  \begin{figure}
 	\caption{Progressive stress-bridging models A through E.}
        \label{fig:bridge2GMC}
  \end{figure}

  \begin{figure}
 	\caption{Stress-bridging paths for Model C.}
        \label{fig:bridge_2C}
  \end{figure}

  \begin{figure}
    \caption{Schematic of the recursive cell method. }
    \label{fig:RCM}
  \end{figure}

  \begin{figure}
    \caption{Comparisons of estimates of $\Ceff{11}$ for (a) models of the dry blend of PBX 9501 (b) models of pressed PBX 9501. }
    \label{fig:dryAll}
  \end{figure}
  \end{flushleft}
  \clearpage
  \begin{table}
      \caption{Component properties used by Greengard and Helsing (1998).}
	\medskip
      \label{tab:helprop}
    \begin{center}
      \begin{tabular}{|l|c|c|c|c|c|}
	\hline
	& Young's & Poisson's & Two-Dimensional & Shear \\
        & Modulus & Ratio     & Bulk Modulus    & Modulus \\
        & (MPa)   &           & (MPa)           & (MPa) \\
	\hline
	Disks   & 324 & 0.20 & 225 & 135 \\
	Binder  & 2.7 & 0.35 & 3.3 &   1 \\
	\hline
      \end{tabular}
    \end{center}
  \end{table}

  \begin{table}
    \caption{Experimentally determined elastic moduli and stiffness of PBX 9501 and its 
             constituents (Wetzel 1999). $C_{ij}$ are components of the stiffness matrix.} \medskip
    \label{tab:exptCPBX}
    \begin{center}
    \begin{tabular}{|l|c|c|c|c|c|}
	\hline
	Material & Young's & Poisson's & $C_{11}$ = $C_{22}$ & $C_{12}$ & $C_{66}$ \\
                 & Modulus & Ratio & & &\\
	            & (MPa) & & (MPa) & (MPa) & (MPa) \\
	\hline
	Particles & 15300 & 0.32 & 21894 & 10303 & 5795 \\
	Binder    & 0.7 & 0.49 & 11.97 & 11.51 & 0.235 \\
	PBX 9501 & 1013 & 0.35 & 1626 & 875 & 375 \\
	\hline
    \end{tabular}
    \end{center}
  \end{table}

  \begin{table}
    \caption{Effective stiffnesses of the six model microstructures
		 from GMC and FEM calculations.} \medskip
    \label{tab:pbxManualGMC}
    \begin{center}
    \begin{tabular}{|l||c|c|c||c|c|c||c|c|c||}
      \hline
      & \multicolumn{3}{c||}{$\Ceff{11}$ (MPa)} & \multicolumn{3}{c||}{$\Ceff{12}$ (MPa)} & 
        \multicolumn{3}{c||}{$\Ceff{66}$ (MPa)} \\
      \cline{2-10}
      & FEM  & \multicolumn{2}{c||}{GMC} & FEM  & \multicolumn{2}{c||}{GMC} & FEM  & 
        \multicolumn{2}{c||}{GMC} \\
      \cline{2-10}
      & & Binary & Effective& & Binary & Effective& & Binary & Effective\\
      & & Subcell & Subcell & & Subcell & Subcell & & Subcell & Subcell \\
      \hline
      Model 1 & 177 & 814 & 479 & 90  & 119 & 103 & 11 & 2.4 & 2.3\\
      Model 2 & 181 & 807 & 477 & 86  & 112 & 103 & 12 & 2.3 & 2.3\\
      Model 3 & 186 & 815 & 193 & 88  & 108 & 89  & 15 & 2.3 & 2.2\\
      Model 4 & 143 & 116 & 142 & 114 & 112 & 124 & 33 & 2.4 & 2.6\\
      Model 5 & 237 & 132 & 323 & 94  & 100 & 104 & 38 & 2.3 & 2.5\\
      Model 6 & 229 & 132 & 334 & 76  & 93  & 100 & 9  & 2.1 & 2.5\\
      \hline
      Mean    & 192 & 471 & 325 & 91  & 107 & 104 & 20 & 2.3 & 2.4\\
      \hline
    \end{tabular}
    \end{center}
  \end{table}

  \begin{table}
    \caption{Effective stiffness of the model PBX 9501 microstructures
		 from GMC and FEM calculations.} \medskip
    \label{tab:pbxFEMGMC}
    \begin{center}
    \begin{tabular}{|l||c|c||c|c||c|c||c|c||}
      \hline
      & \multicolumn{2}{c||}{$\Ceff{11}$ (MPa)} & \multicolumn{2}{c||}{$\Ceff{22}$ (MPa)} & 
        \multicolumn{2}{c||}{$\Ceff{12}$ (MPa)} & \multicolumn{2}{c||}{$\Ceff{66}$ (MPa)} \\
      \cline{2-9}
      & FEM  & GMC & FEM  & GMC & FEM  & GMC & FEM & GMC\\
      \hline
      Model DB1 & 2385 & 152 & 2094 & 148 & 633  & 122 & 792 & 4.9 \\
      Model DB2 & 3618 & 146 & 1643 & 144 & 656  & 122 & 750 & 4.9 \\
      Model DB3 & 3546 & 149 & 3385 & 148 & 1142 & 125 & 1317 & 5.0 \\
      Model DB4 & 5274 & 144 & 5124 & 146 & 1712 & 120 & 1703 & 4.8 \\
      \hline
      Model PP1 & 3180 & 180 & 3570 & 188 & 989  & 131 & 1262 & 4.8 \\
      Model PP2 & 3886 & 170 & 3683 & 190 & 1032 & 132 & 1278 & 5.7 \\
      Model PP3 & 6302 & 181 & 6221 & 181 & 2043 & 133 & 2077 & 6.6 \\
      Model PP4 & 7347 & 182 & 7587 & 186 & 2547 & 129 & 2542 & 6.9 \\
      \hline
    \end{tabular}
    \end{center}
  \end{table}

  \begin{table}
      \caption{The elastic properties of the components of the X shaped
	       microstructure. $C_{ij}$ are components of the stiffness matrix.} \medskip
      \label{tab:bridge1atoe}
    \begin{center}
      \begin{tabular}{|l|c|c|c|c|c|}
	\hline
	& Young's & Poisson's & $C_{11}$ & $C_{12}$ & $C_{66}$ \\
        & Modulus & Ratio & & &\\
	      & (MPa) &  & (MPa) & (MPa) & (MPa) \\
	\hline
	Particles & 15300 & 0.32 & 21894 & 10303 & 5795 \\
        Binder a & 0.7  & 0.49 & 12   & 11.5   & 0.2 \\
        Binder b & 7    & 0.49 & 120    & 115  & 2.4  \\
        Binder c & 70   & 0.49 & 1198   & 1151   & 23.5   \\
        Binder d & 700  & 0.49 & 11980  & 11510  & 235  \\
        Binder e & 7000 & 0.49 & 119799 & 115101 & 2349 \\
	\hline
      \end{tabular}
    \end{center}
  \end{table}

  \begin{table}
    \caption{Effective properties of edge bridging models.} \medskip
    \label{tab:bridge2GMC}
    \begin{center}
    \begin{tabular}{|l||c|c||c|c||c|c||c|c||}
    \hline
    & \multicolumn{2}{c||}{\Ceff{11} (MPa)} & \multicolumn{2}{c||}{\Ceff{22} (MPa)}
    & \multicolumn{2}{c||}{\Ceff{12} (MPa)} & \multicolumn{2}{c||}{\Ceff{66} (MPa)} \\
    \cline{2-9}
    & FEM & GMC & FEM & GMC & FEM & GMC & FEM & GMC \\
    \hline
    Model A & 16    & 16   & 16    & 16   & 15   & 15   & 0.4  & 0.3 \\
    Model B & 336   & 19   & 343   & 19   & 337  & 18   & 537  & 0.4 \\
    Model C & 4095  & 25   & 889   & 24   & 1470 & 23   & 1093 & 0.5 \\
    Model D & 8992  & 8540 & 1361  & 32   & 523  & 23   & 1182 & 0.6 \\
    Model E & 10017 & 9042 & 10052 & 9042 & 2892 & 2143 & 1799 & 0.9 \\
    \hline
    \end{tabular}
    \end{center}
  \end{table}
  \clearpage
     \begin{center}
	\scalebox{0.4}{\includegraphics{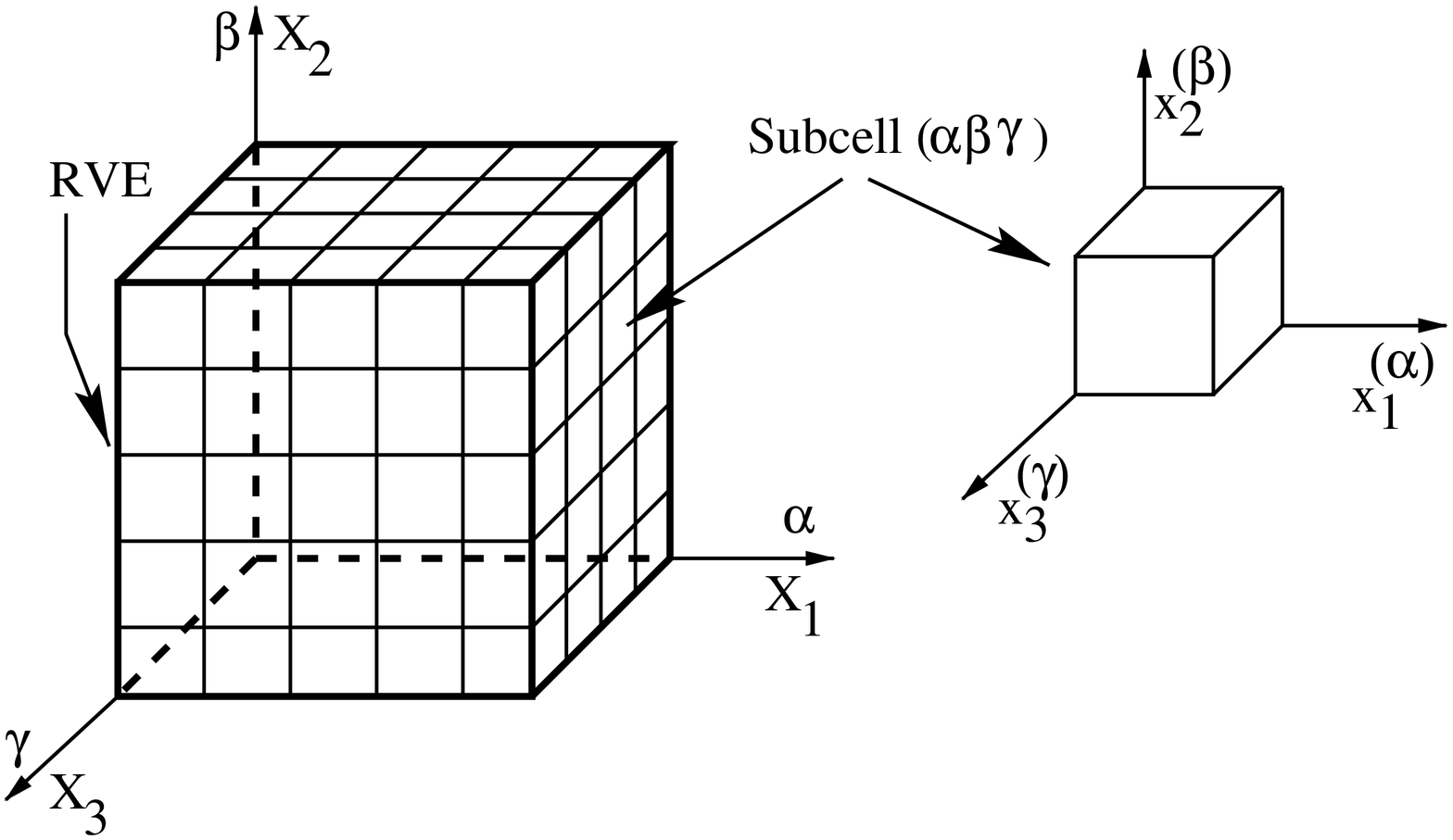}}
        \\ \vspace{24pt} Figure 1 
     \end{center}
   \newpage

    \begin{center}
      \scalebox{0.52}{\includegraphics{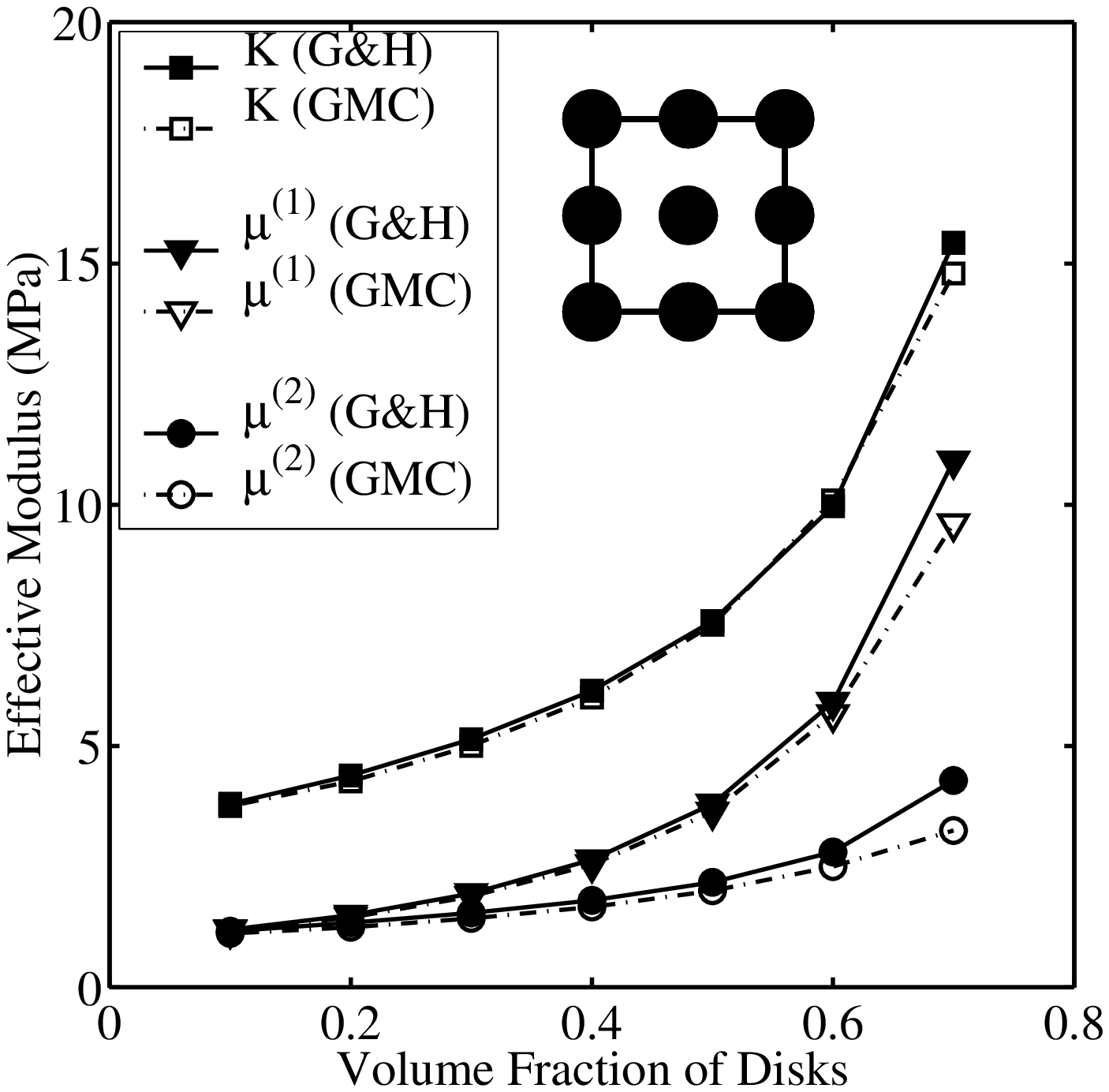}}
        \\ \vspace{24pt} Figure 2 
    \end{center}
   \newpage

     \begin{center}
        \scalebox{0.65}{\includegraphics{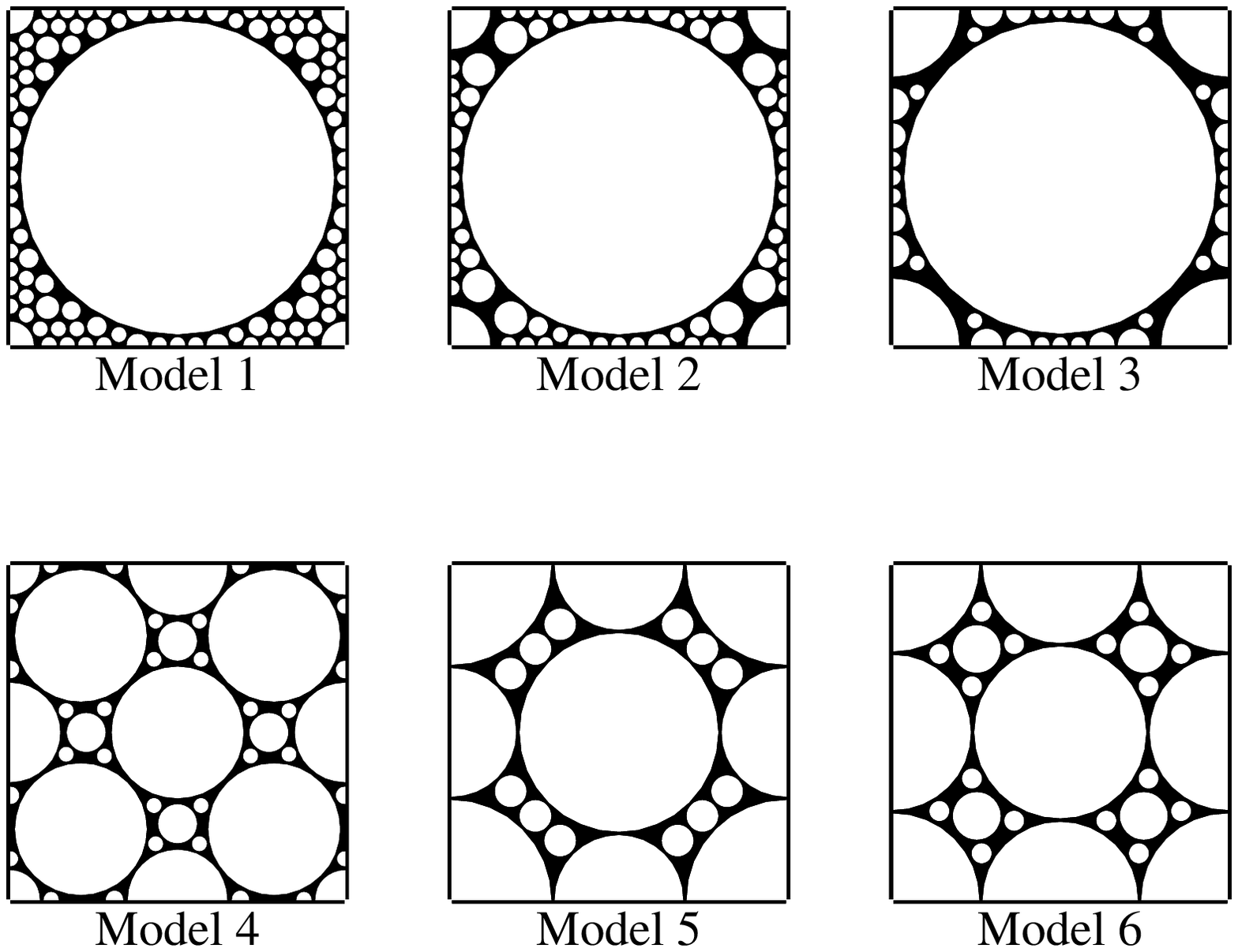}}
        \\ \vspace{24pt} Figure 3 
     \end{center}
   \newpage

     \begin{center}
	\scalebox{0.4}{\includegraphics{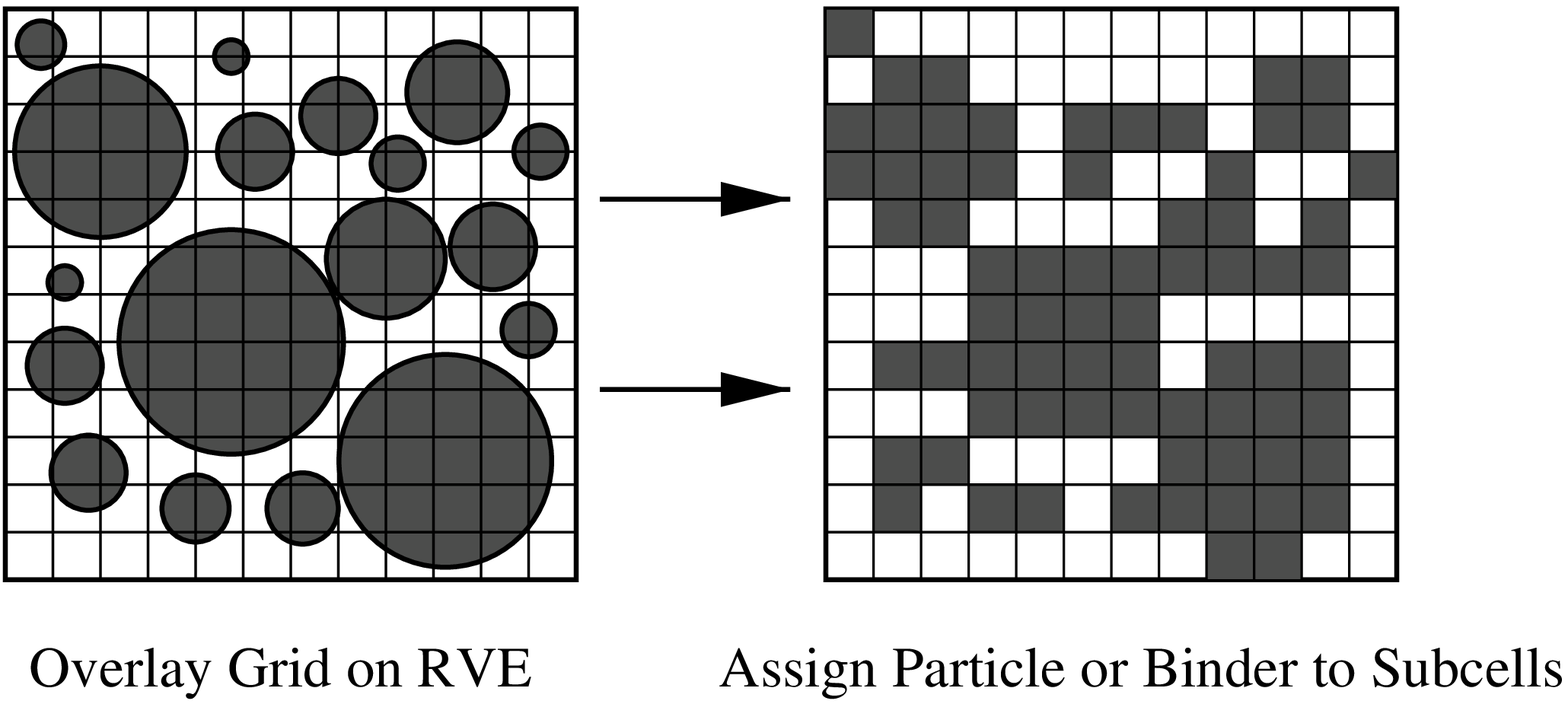}} \\
        (a) Binary subcell approach.\\\vspace{15pt}
	\scalebox{0.4}{\includegraphics{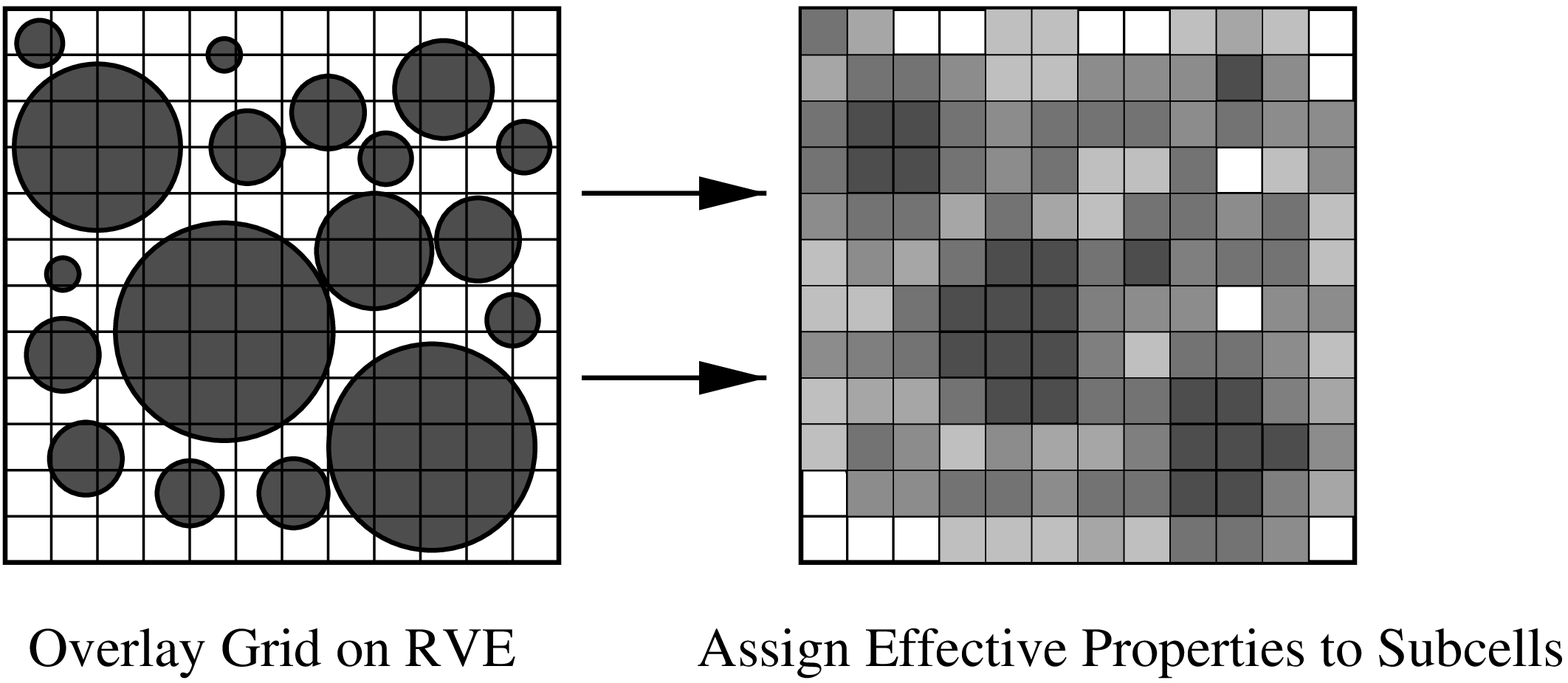}} \\
	(b) Effective subcell approach.
        \\ \vspace{24pt} Figure 4 
     \end{center}
   \newpage

     \begin{center}
        \scalebox{0.24}{\includegraphics{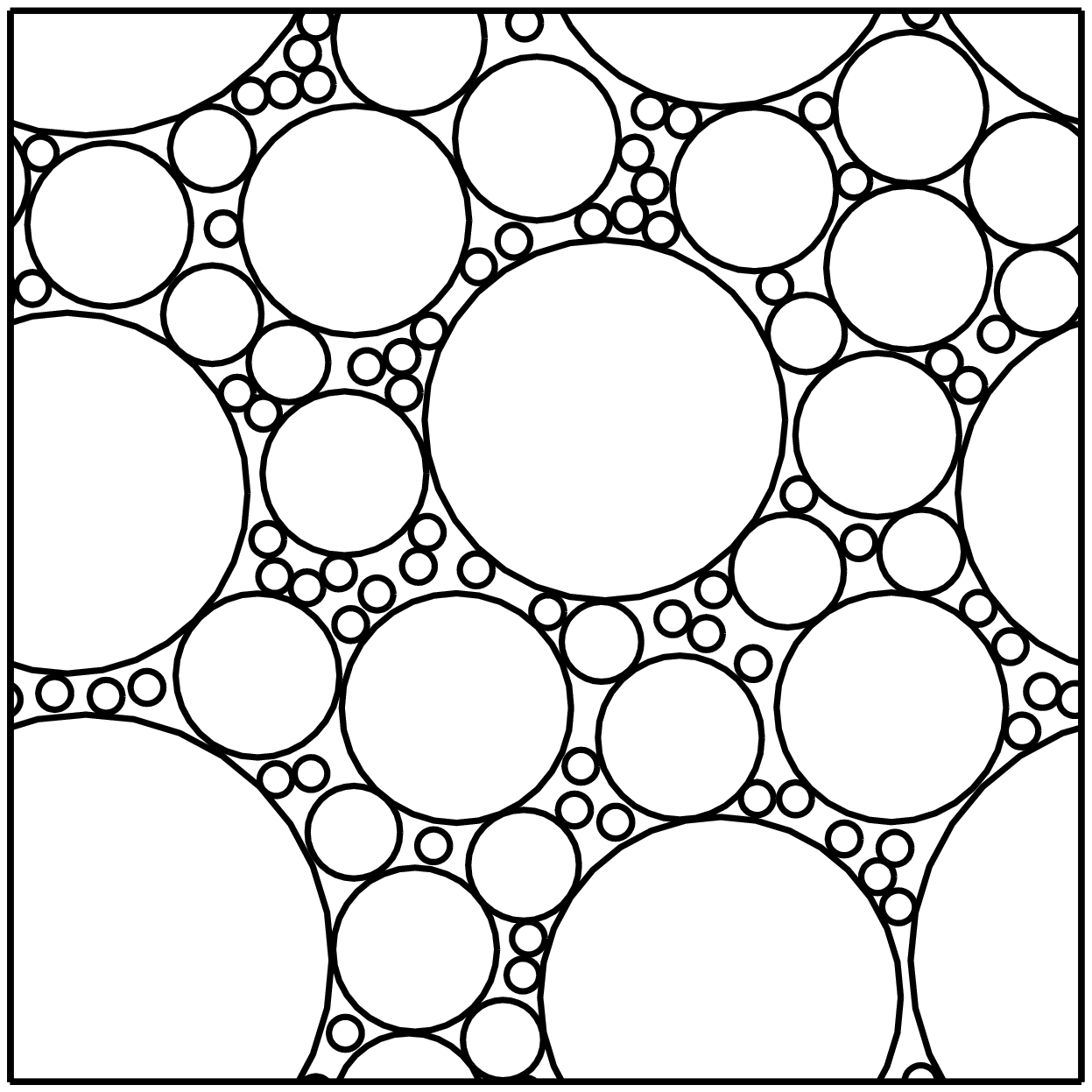}\hspace{48pt}
                       \includegraphics{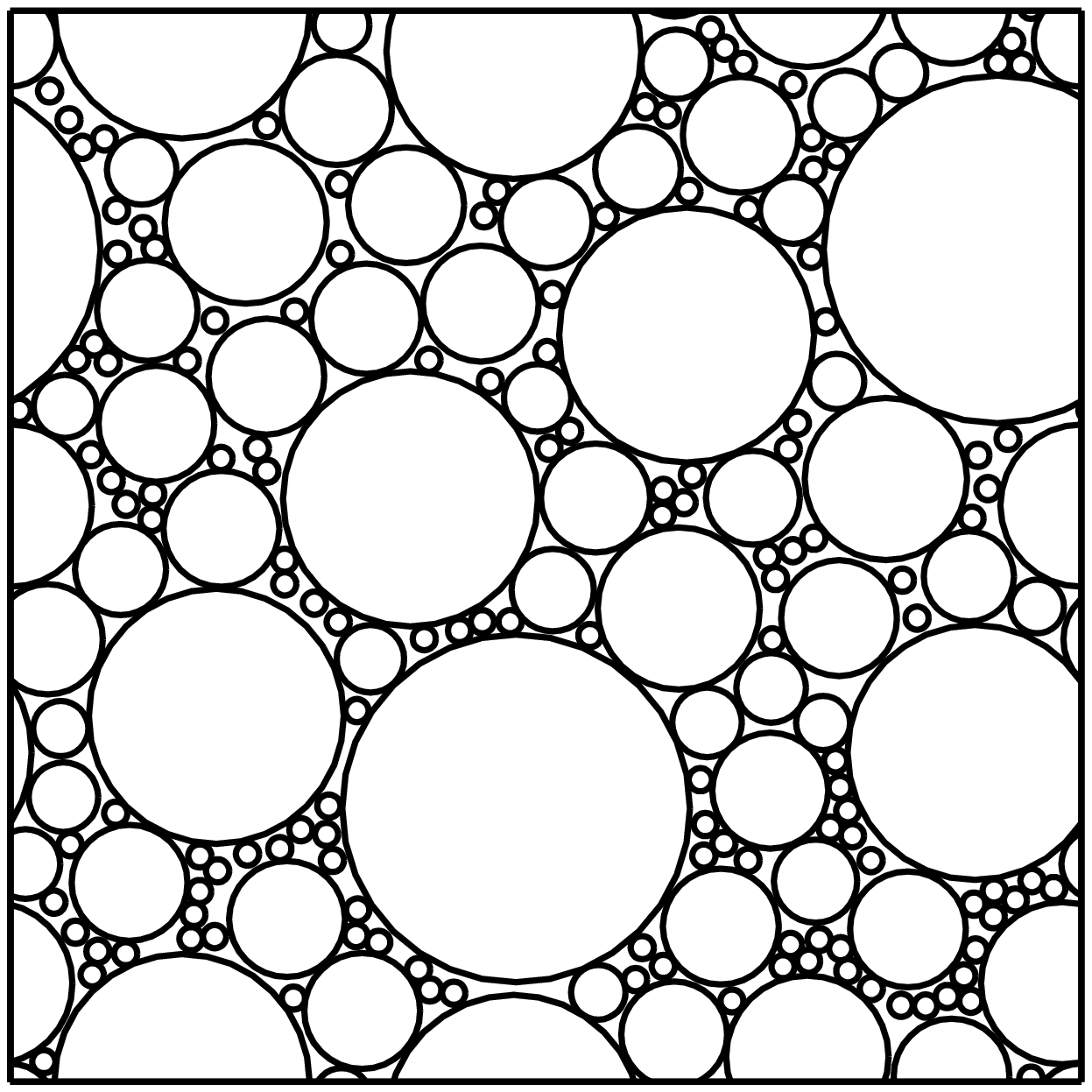}\hspace{48pt}
                       \includegraphics{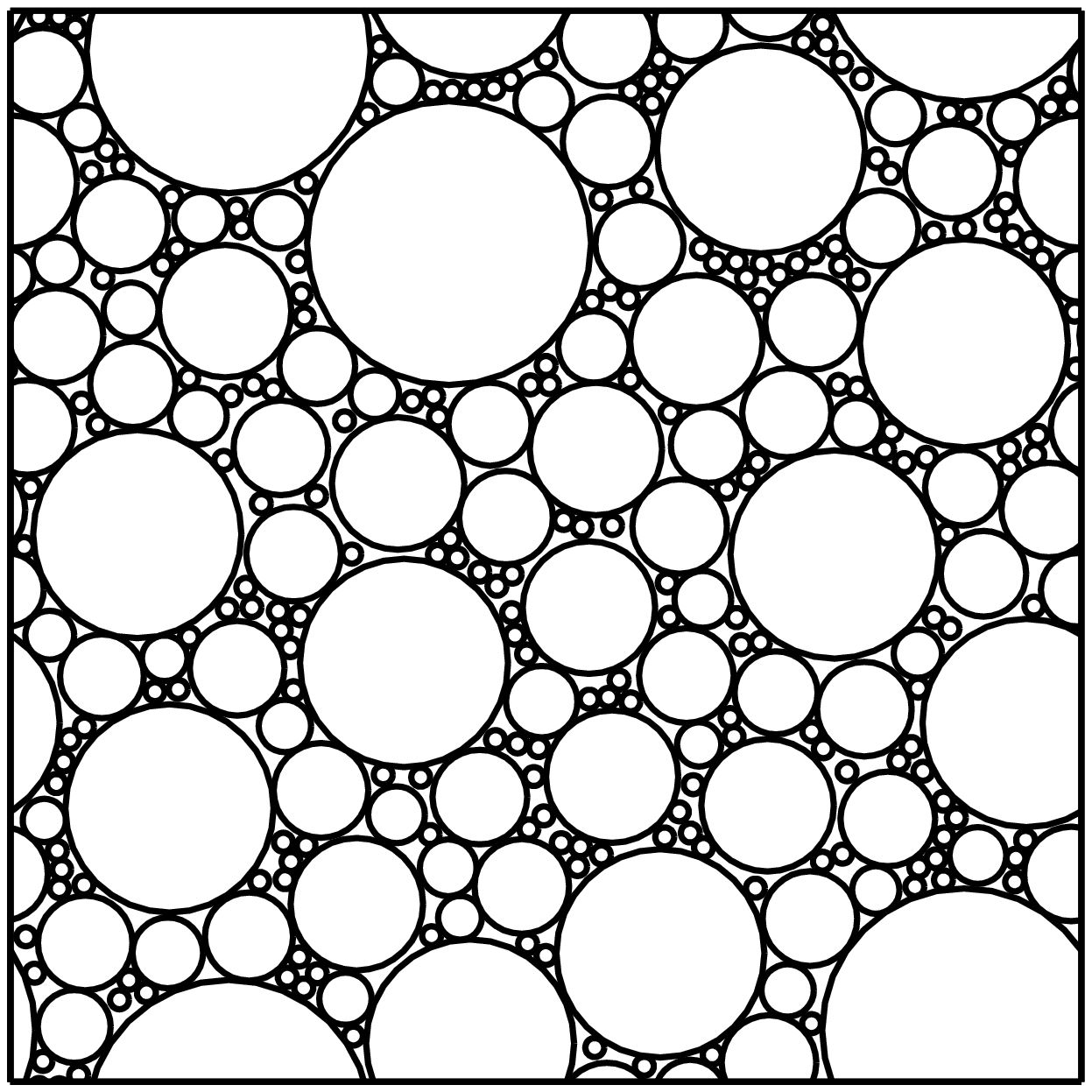}\hspace{48pt}
                       \includegraphics{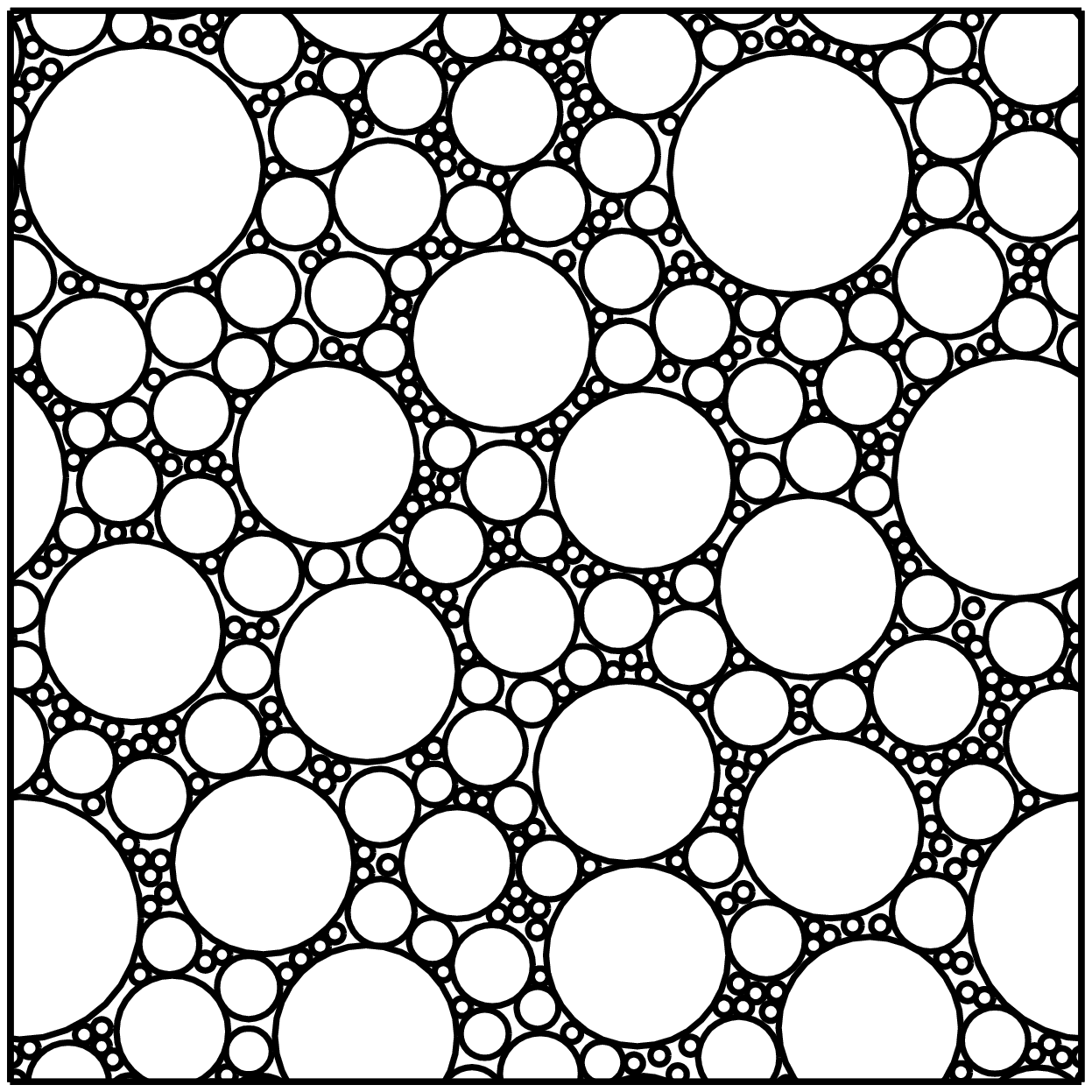}}\\
        0.65$\times$0.65 mm$^2$ \hspace{12pt}
        0.94$\times$0.94 mm$^2$ \hspace{12pt}
        1.13$\times$1.13 mm$^2$ \hspace{12pt}
        1.33$\times$1.33 mm$^2$ \\
        DB1\hspace{1.2in}DB2\hspace{1.2in}DB3\hspace{1.2in}DB4\\
        (a) Dry Blend of PBX 9501 \\\vspace{24pt}
        \scalebox{0.24}{\includegraphics{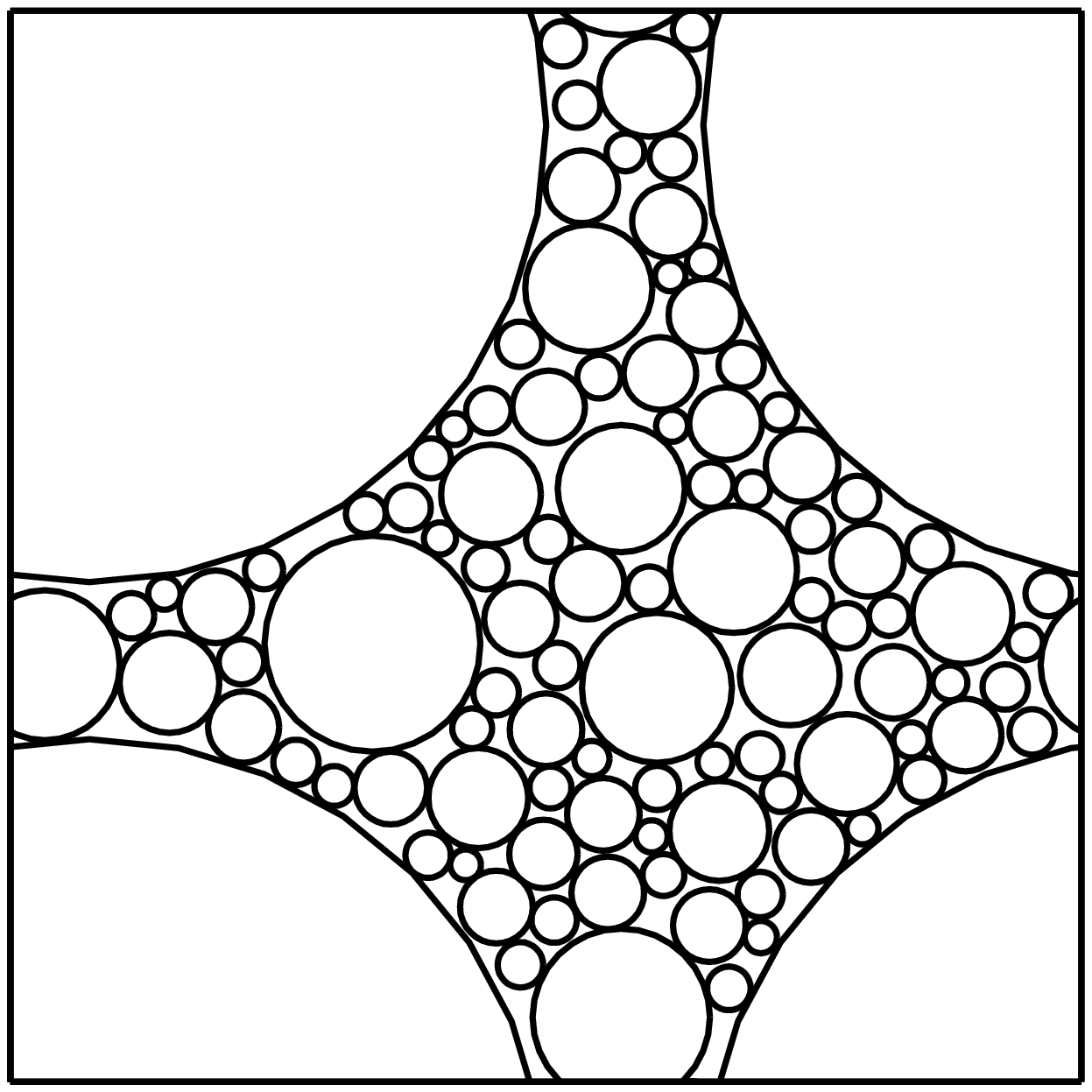}\hspace{48pt}
                       \includegraphics{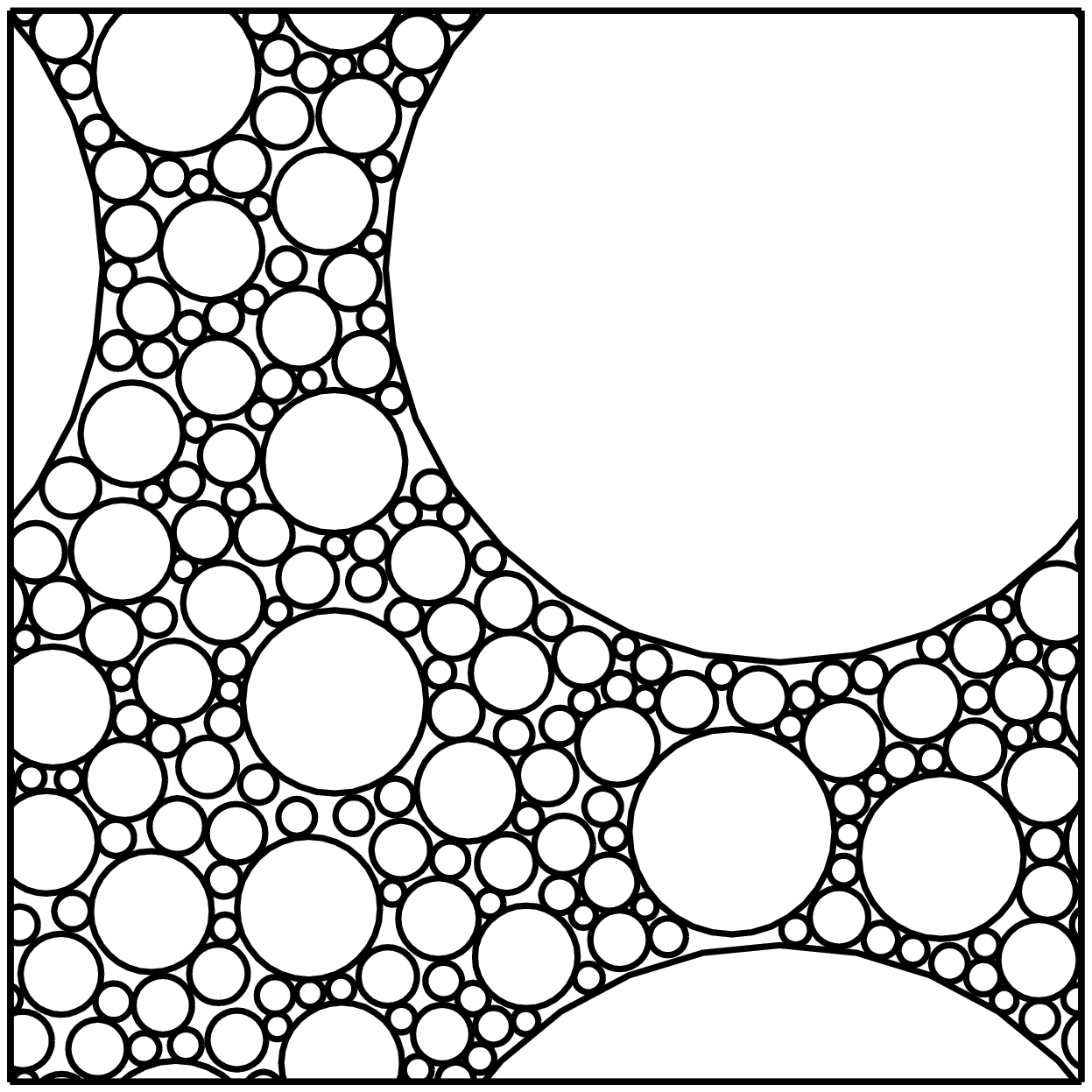}\hspace{48pt}
                       \includegraphics{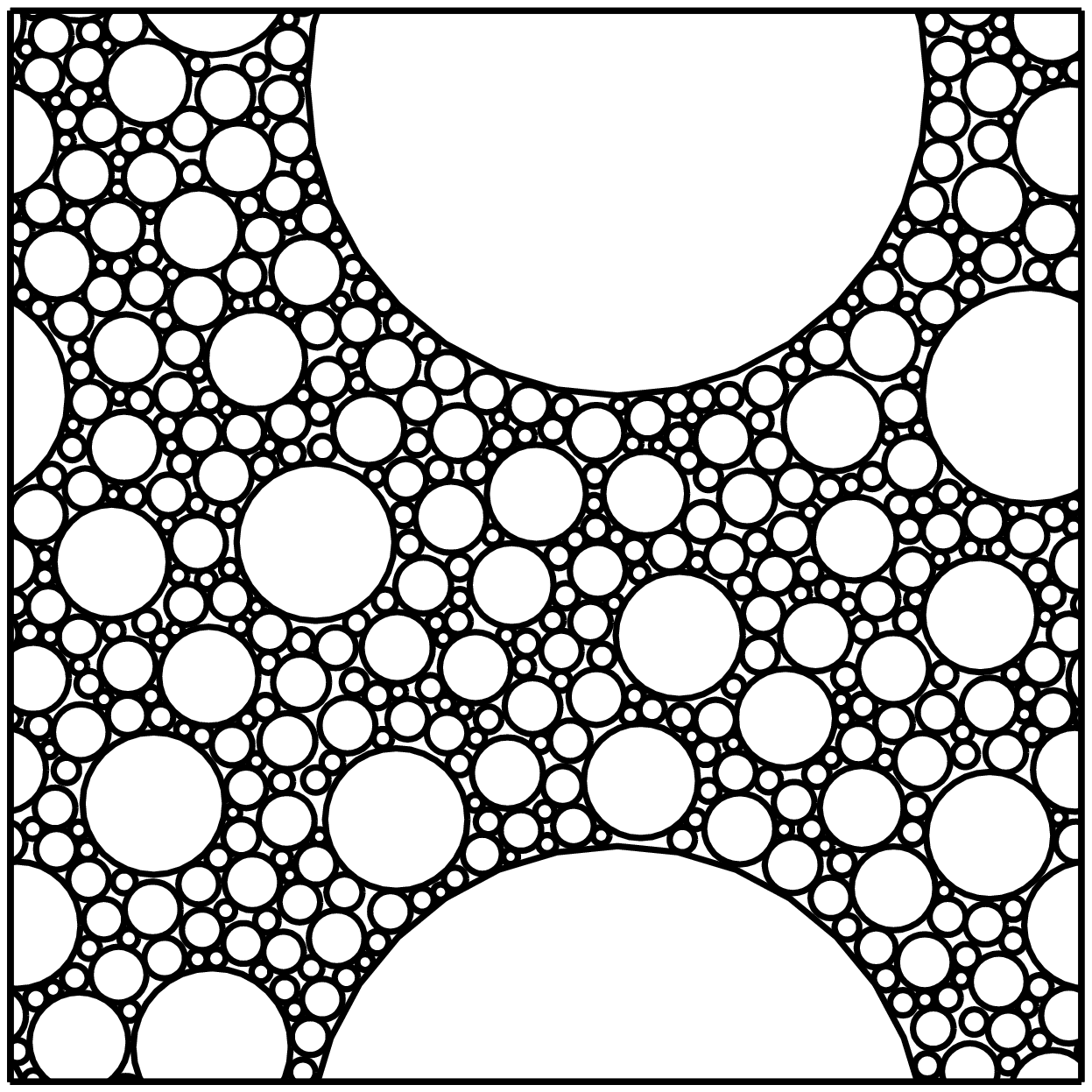}\hspace{48pt}
                       \includegraphics{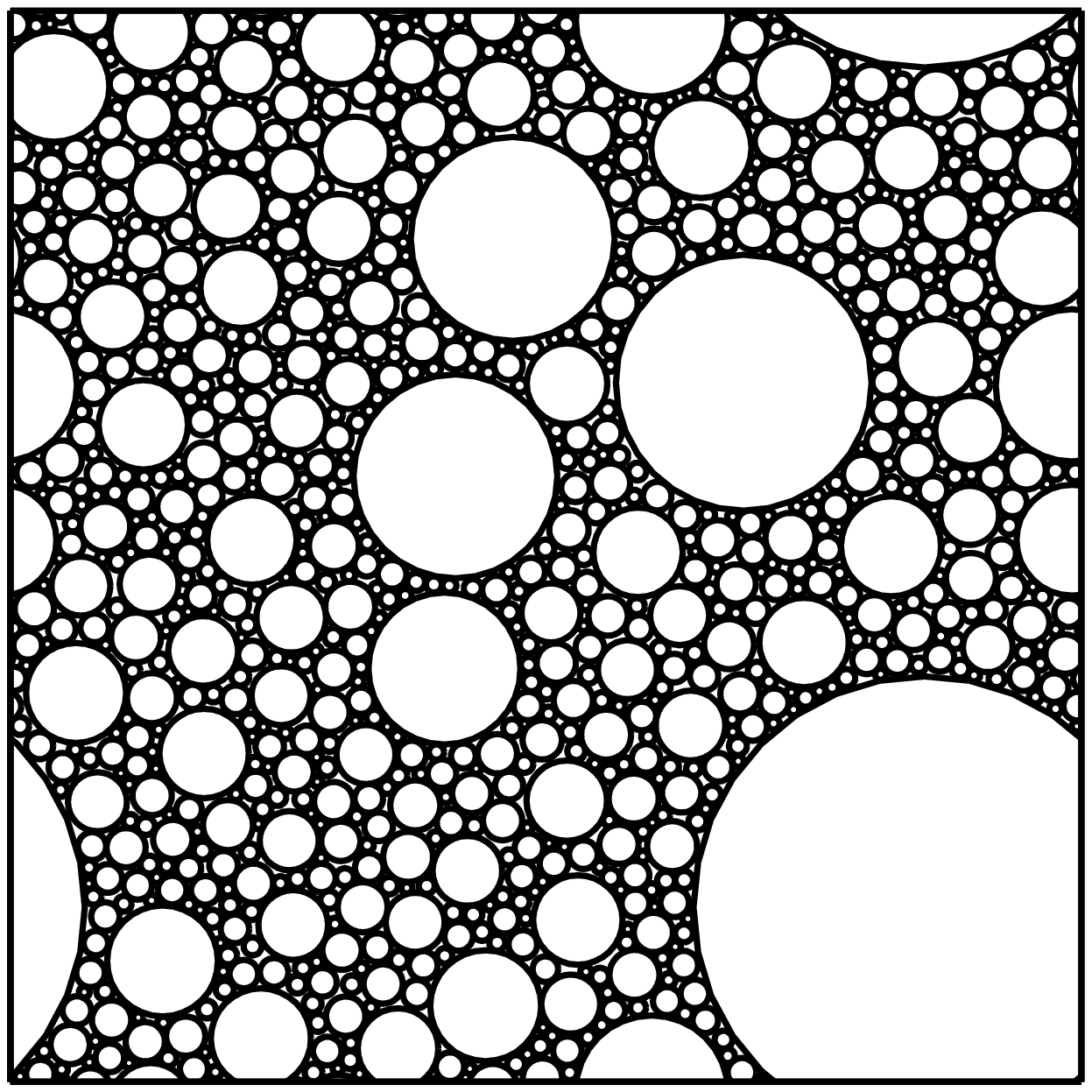}}\\
        0.36$\times$0.36 mm$^2$ \hspace{12pt}
        0.42$\times$0.42 mm$^2$ \hspace{12pt}
        0.54$\times$0.54 mm$^2$ \hspace{12pt}
        0.68$\times$0.68 mm$^2$ \\
        PP1\hspace{1.2in}PP2\hspace{1.2in}PP3\hspace{1.2in}PP4\\
        (b) Pressed PBX 9501
        \\ \vspace{24pt} Figure 5 
     \end{center}
   \newpage

     \begin{center}
	\scalebox{0.6}{\includegraphics{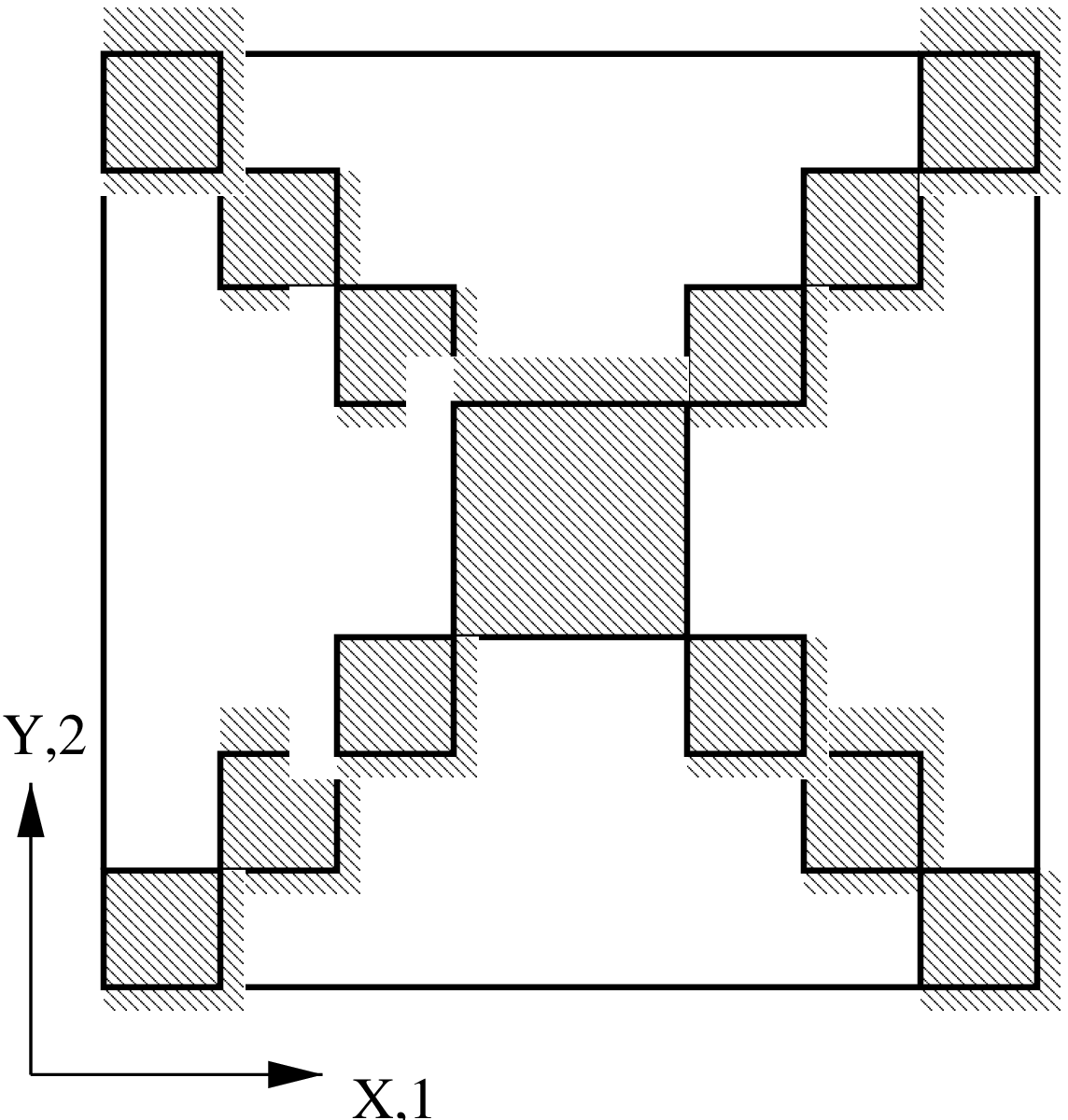}}
        \\ \vspace{24pt} Figure 6 
     \end{center}
   \newpage

     \begin{center}
	\scalebox{0.55}{\includegraphics{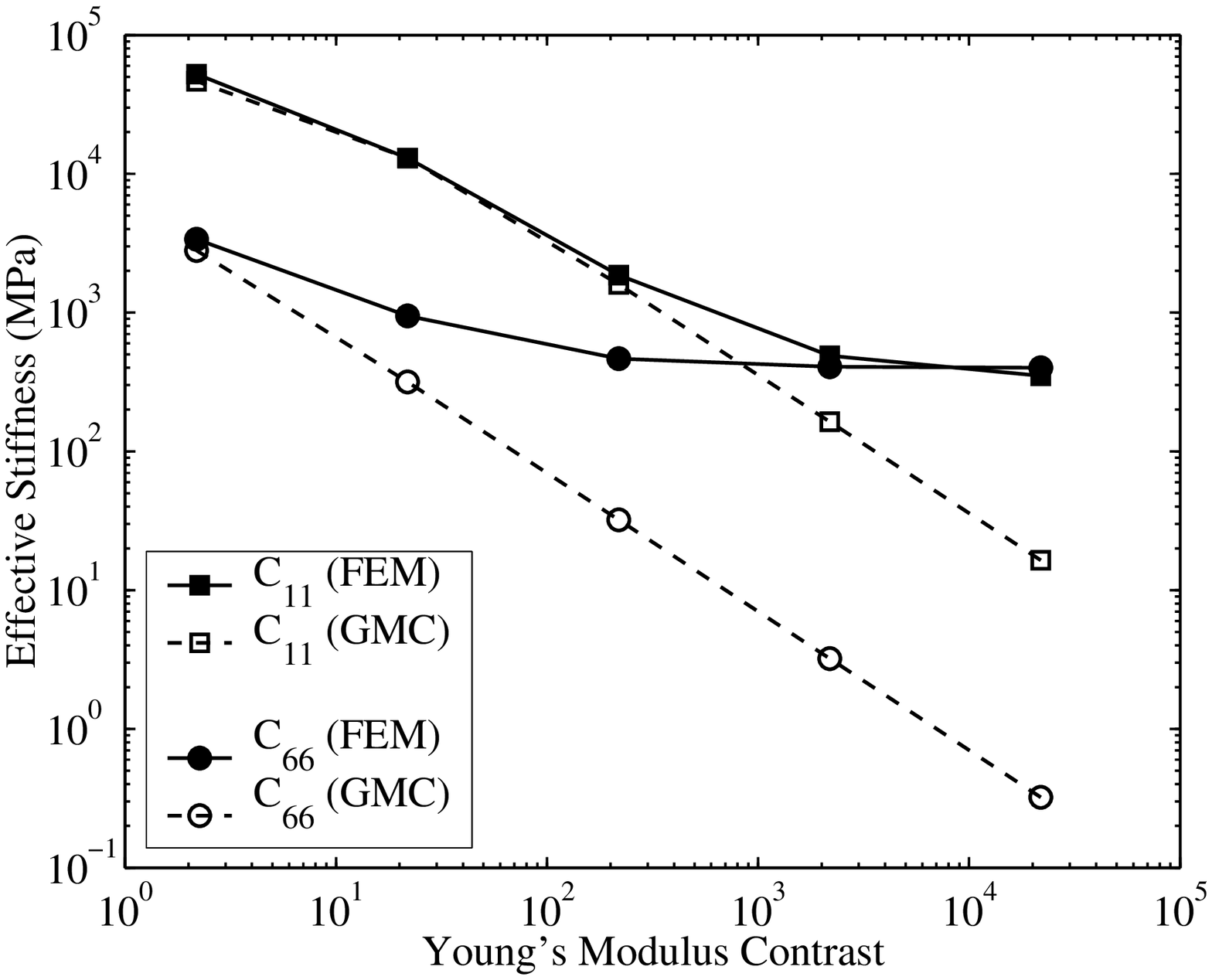}}
        \\ \vspace{24pt} Figure 7 
     \end{center}
   \newpage

     \begin{center}
	\scalebox{0.3}{\includegraphics{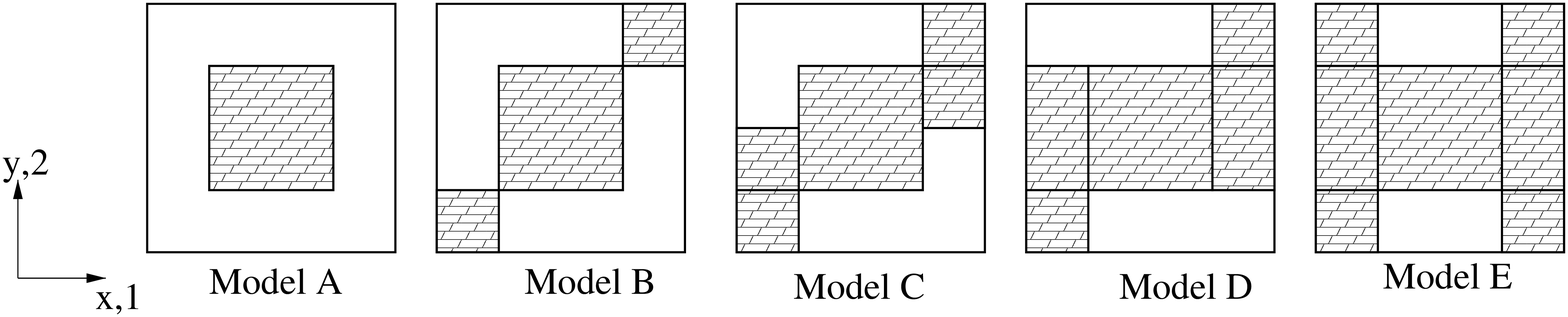}}
        \\ \vspace{24pt} Figure 8 
     \end{center}
   \newpage

     \begin{center}
	\scalebox{0.4}{\includegraphics{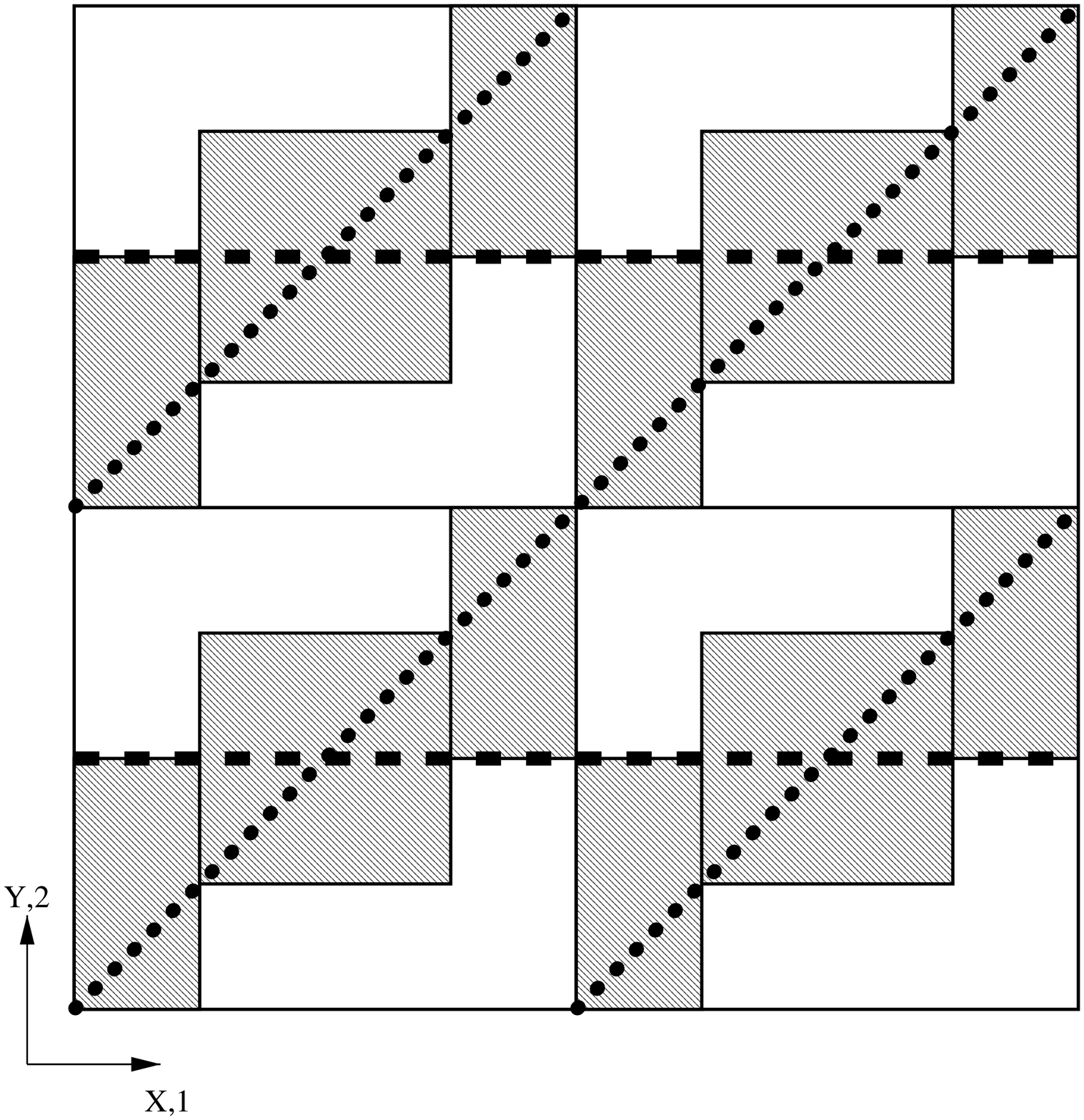}}
        \\ \vspace{24pt} Figure 9 
     \end{center}
   \newpage

    \begin{center}
    \scalebox{0.5}{\includegraphics{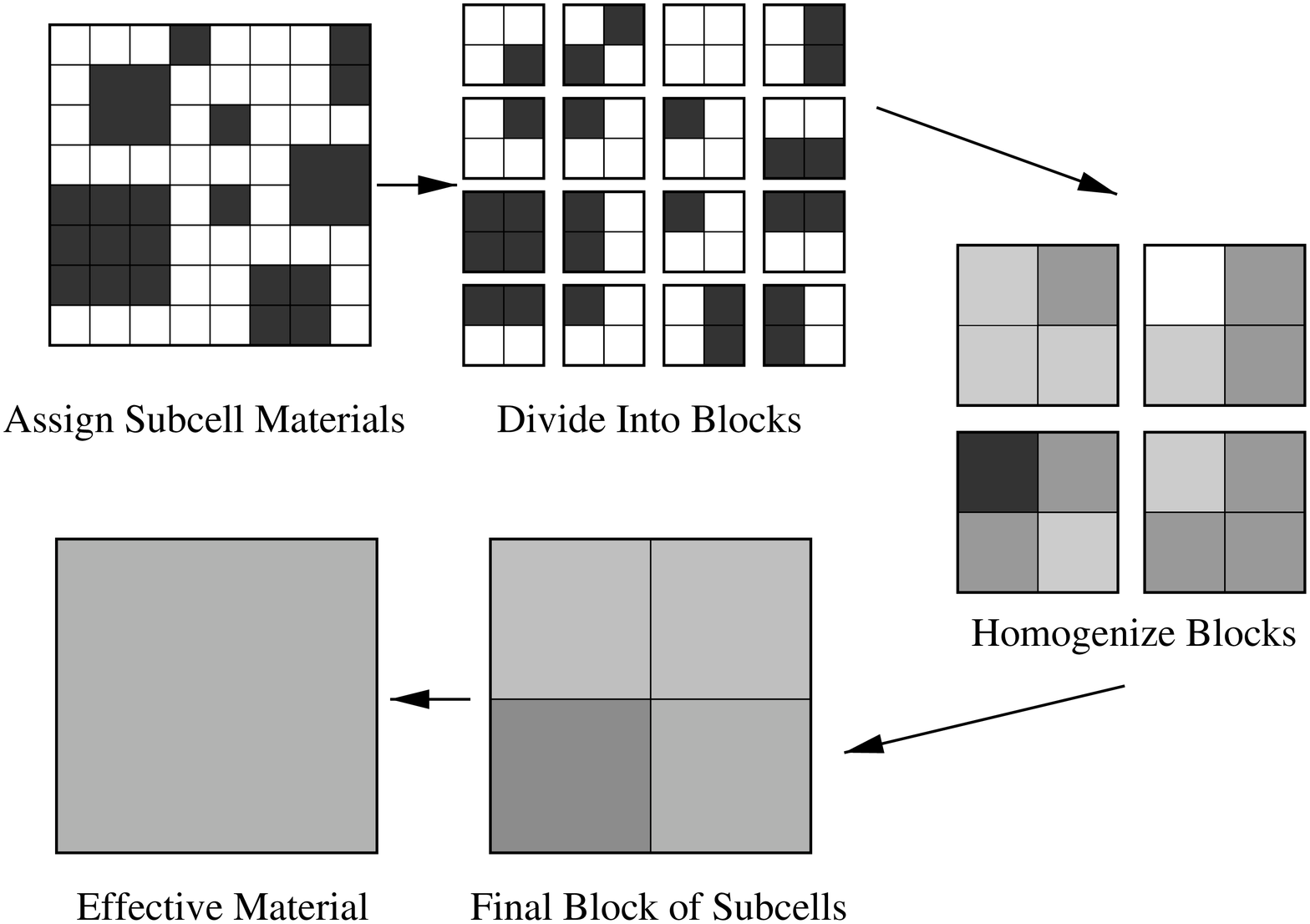}}
        \\ \vspace{24pt} Figure 10 
    \end{center}
   \newpage

    \begin{center}
    \scalebox{0.5}{\includegraphics{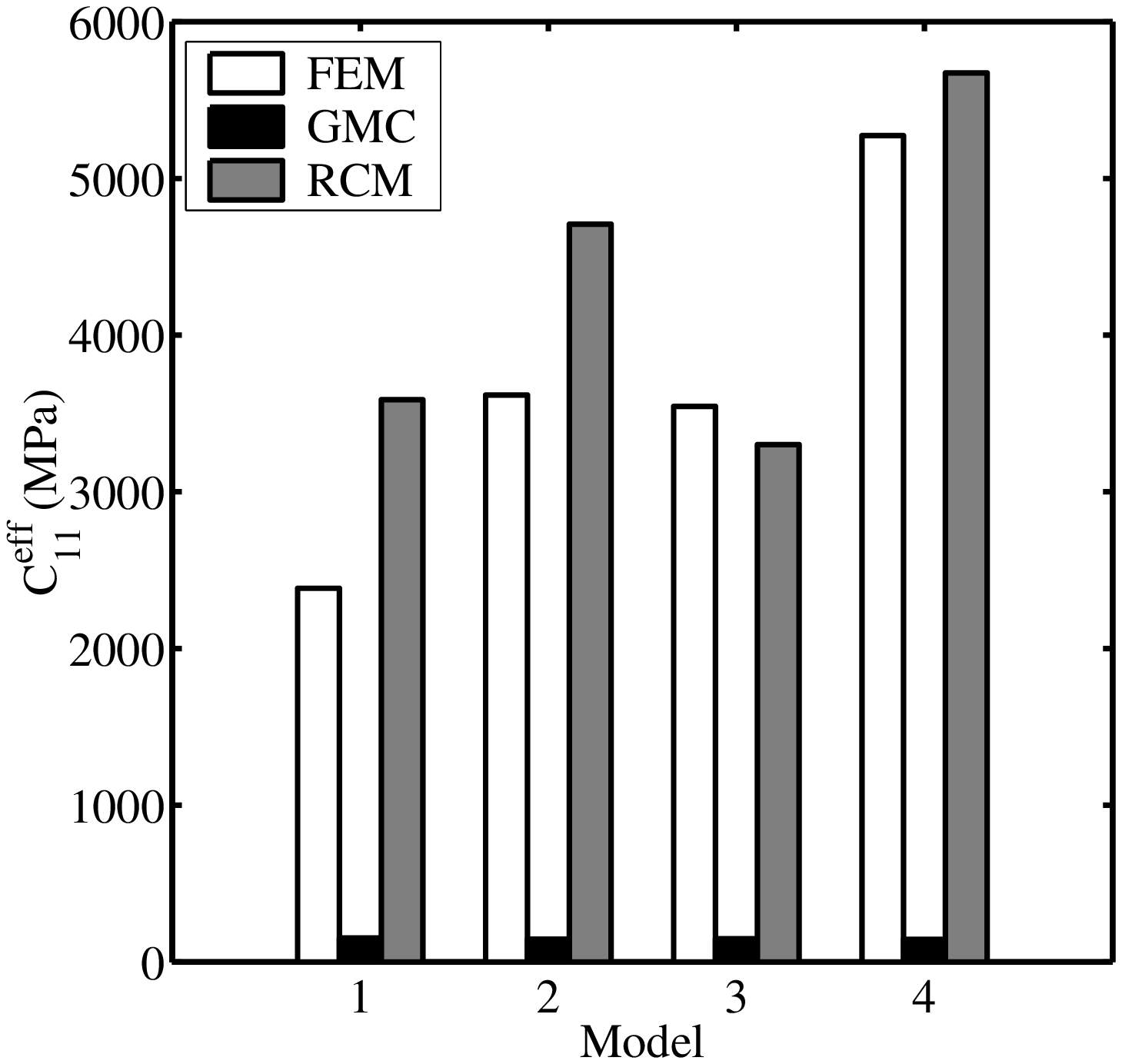}\includegraphics{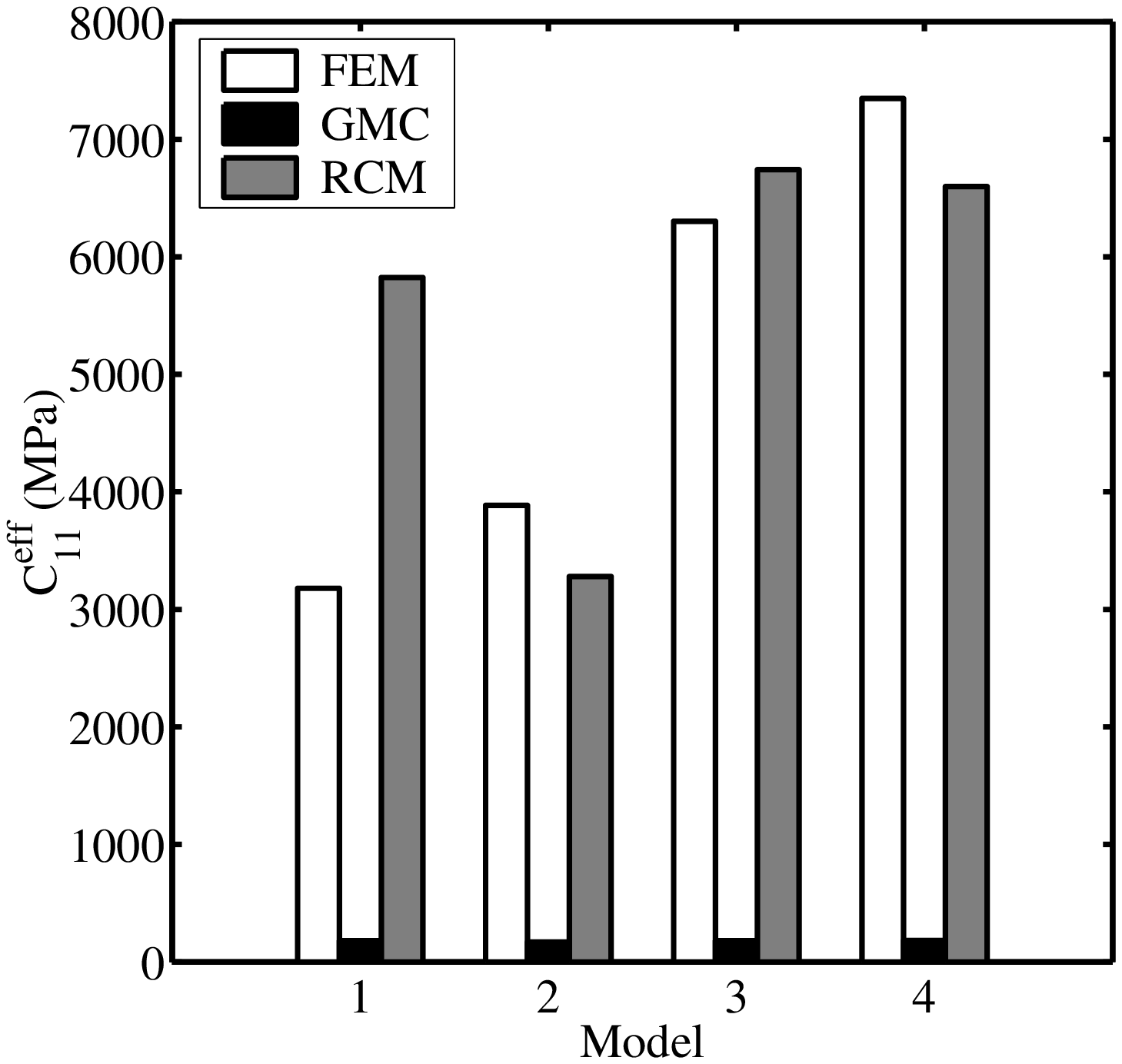}}\\
    (a) Dry Blend \hspace{1.5in} (b) Pressed PBX
        \\ \vspace{24pt} Figure 11 
    \end{center}

\end{document}